\documentclass[journal=cmatex,manuscript=article]{achemso}

\usepackage{graphicx}
\usepackage{amsmath}
\usepackage{amssymb}
\usepackage{booktabs}
\usepackage{multirow}

\author{Lior Verbitsky}
\affiliation{Department of Chemistry, Princeton University, Princeton, New Jersey 08544, USA}
\author{Amira Merino}
\affiliation{Department of Chemistry, Princeton University, Princeton, New Jersey 08544, USA}
\author{Scott B. Lee}
\affiliation{Department of Chemistry, Princeton University, Princeton, New Jersey 08544, USA}
\author{Jaime M. Moya}
\affiliation{Department of Chemistry, Princeton University, Princeton, New Jersey 08544, USA}
\author{Sigalit Aharon}
\affiliation{Department of Chemistry, Princeton University, Princeton, New Jersey 08544, USA}
\author{Fatmag\"ul Katmer}
\affiliation{Department of Chemistry, Princeton University, Princeton, New Jersey 08544, USA}
\author{Sudipta Chatterjee}
\affiliation{Department of Chemistry, Princeton University, Princeton, New Jersey 08544, USA}
\author{Grigorii Skorupskii}
\affiliation{Department of Chemistry, Princeton University, Princeton, New Jersey 08544, USA}
\author{Josh Leeman}
\affiliation{Department of Chemistry, Princeton University, Princeton, New Jersey 08544, USA}
\author{Gabrielle Carrel}
\affiliation{Department of Chemistry, Princeton University, Princeton, New Jersey 08544, USA}
\author{Leslie M. Schoop}
\email{lschoop@princeton.edu}
\affiliation{Department of Chemistry, Princeton University, Princeton, New Jersey 08544, USA}

\title{New Superconductors in the PtPb$_3$Bi Structure Type}

\begin{document}

% ---------------------------------------------------------------------
\begin{abstract}
The quest for new superconductors is of both fundamental and technological importance. Recently, an artificial intelligence method correctly predicted PtPb$_3$Bi to be a superconductor. In this work, we find superconductivity in the newly synthesized $M$Pb$_{4-x}$Bi$_x$ ($M$~=~Au, Pd, and Rh), of which PtPb$_3$Bi is a member. When
$M$~=~Ni, whose radius is considerably smaller, the structure instead
collapses into the different, Pb-substituted NiBi$_3$ type. Interestingly, the stoichiometric parameter $x$ shifts across the three compounds to keep the total valence electron count close to 20 per formula unit. The superconducting transitions occur at 4.9, 4.2, and 3.4~K, for $M$~=~Au, Pd, and Rh, respectively.  Using electrical resistivity, magnetization, and specific heat measurements, we establish the bulk nature of the superconducting
state and determine the critical fields, characteristic length scales,
and anisotropy ratios. All three compounds are moderately anisotropic
type-II superconductors, with modest upper critical field anisotropies
of $H_{c2}^{\parallel c}/H_{c2}^{\perp c}\approx1.2$ to $1.5$. These results
establish $M$Pb$_{4-x}$Bi$_x$ as a family of anisotropic
superconductors and a platform for studying how site disorder and
Pb-Bi mixing govern superconductivity in heavy-element intermetallics.

\end{abstract}

\maketitle

% =====================================================================
\section{Introduction}
% =====================================================================
One of the intriguing features of superconductivity is that it often
recurs within a given structure type, persisting robustly across
substitutions and distortions. Prominent examples include the A15
compounds,\cite{stewart2015superconductivity} the Chevrel
phases,\cite{chevrel1986superconducting} the cuprate high-$T_c$
superconductors,\cite{armitage2010progress,chu2015hole} and the
iron-based superconductors. The last remain superconducting not
only across a wide range of related structure
types\cite{paglione2010high,chen2014iron} but also in structurally
related, iron-free analogues.\cite{schoop2013superconductivity}
Identifying a new structure type therefore opens the door to systematic
studies of structure-property relationships, in the spirit of the early
empirical rules formulated by Matthias.\cite{matthias1957chapter} Yet,
even though many structural families have been
discovered,\cite{gui2021chemistry} together amounting to thousands of
superconductors,\cite{yao2021superconducting} predicting the
superconducting transition temperature $T_c$ of a given compound, let
alone designing one with a higher $T_c$, remains difficult and contested.

\begin{figure*}[ht]
\includegraphics{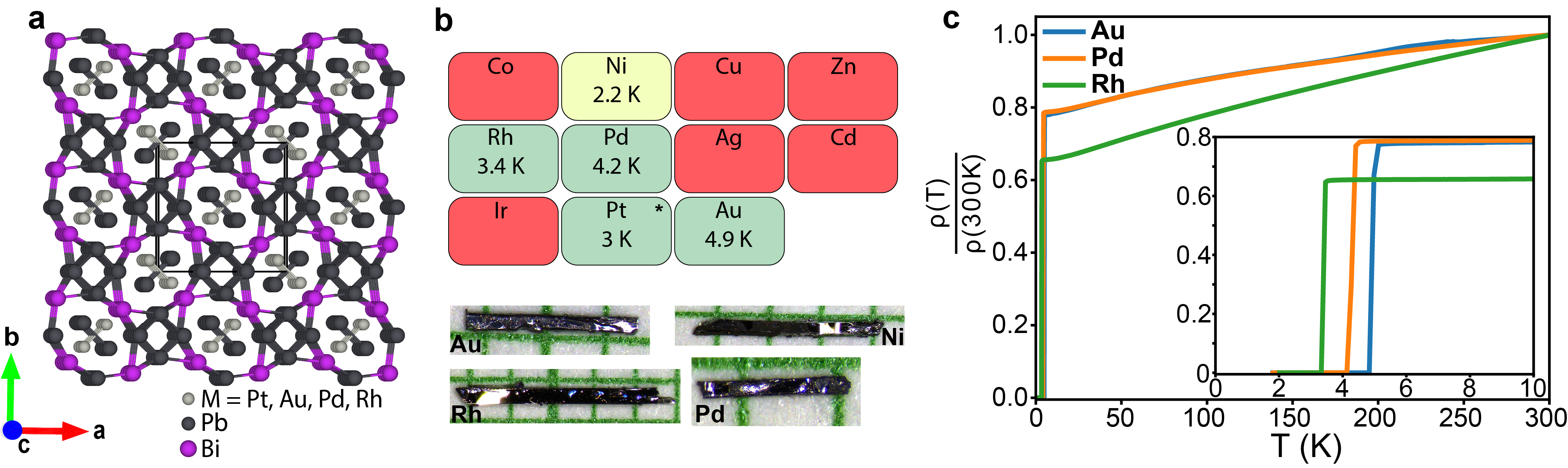}

\caption{\label{fig:fig1} (a) Idealized crystal structure of the $M$Pb$_{4-x}$Bi$_x$
family, based on the archetypal PtPb$_3$Bi structure. Gray, black, and purple spheres denote $M$, Pb, and Bi atoms,
respectively, and the black rectangle outlines the unit cell. (b) Summary
of the $M$ substitution survey. Green cells mark successful synthesis of the
$M$Pb$_{4-x}$Bi$_x$ structure together with the measured $T_c$; the yellow cell
marks the structurally related NiBi$_3$ type; red cells mark attempts
that did not yield a related structure. The asterisk on the Pt cell
denotes the previously reported
parent.\cite{lesser2025learning,srivastava2026discovery} Representative
crystals on millimeter graph paper are shown below the grid. (c)
Normalized resistivity for $M$~=~Au (blue), Pd (orange), and Rh (green)
between 300 and 1.8~K. The inset highlights the sharp drop to zero
resistivity at low temperature.}
\end{figure*}

Because a first-principles calculation of $T_c$ from structure remains
out of reach for most materials, recent work has turned to machine
learning to identify structural descriptors that correlate with
superconductivity. One such descriptor, the distribution of
electron-affinity differences between neighboring atoms, was recently
shown to have predictive power for $T_c$ and guided our discovery of
superconductivity in PtPb$_3$Bi.\cite{lesser2025learning} Until now,
PtPb$_3$Bi was the only known superconductor of its structure
type (ICSD\cite{ICSD} numbers 58834,\cite{matkovic1978kristallstruktur} 58835,\cite{biswas1969strukturuntersuchungen} 197311\cite{schubert1968einige}). The structure is tetragonal and is
built from a Pb-Bi framework that encloses octagonal channels; within
each channel, mutually perpendicular Pt-Pb dimers stack in the $ab$ plane
(Figure~\ref{fig:fig1}a).

Based on the general idea of superconductivity ``running in structure types'', and with, to the best of our knowledge, PtPb$_3$Bi being the only structure of its type, we set out to expand the family by substituting other transition metals at the $M$ site, while preserving the Pb-Bi framework. This
straightforward strategy proved fruitful. We obtained three new
superconductors, $M$Pb$_{4-x}$Bi$_x$ with $M$~=~Au, Pd, and Rh, all
isostructural with PtPb$_3$Bi. Because the measured Pb:Bi ratio deviates
from the ideal 3:1 value found in PtPb$_3$Bi, we denote these compounds
$M$Pb$_{4-x}$Bi$_x$ throughout. When $M$~=~Ni, whose radius is
considerably smaller, the octagonal channels collapse into hexagonal ones
and the material instead adopts the related NiBi$_3$ structure type,
which is itself a superconductor.\cite{silva2013superconductivity}
Substitution with several other transition metals did not yield related compounds under the conditions explored. Taken together, these results
establish $M$Pb$_{4-x}$Bi$_x$ as a small family of anisotropic, type-II
superconductors and a promising platform for studying how site disorder
and Pb-Bi mixing influence superconducting properties.

% =====================================================================
\section{Experimental Section}
% =====================================================================
\paragraph{Crystal growth.}
Single crystals of $M$Pb$_{4-x}$Bi$_x$ were grown by a Pb-Bi alloy self-flux
reaction in a Canfield crucible set, using a 3:2 atomic ratio of Pb to Bi
as the flux. The atomic fraction of $M$ in the total charge was 11\%,
5.5\%, 2\%, and 11\% for Au, Pd, Rh, and Ni, respectively. The nominal atomic composition of the mixtures used was Au$_{11}$Pb$_{53.4}$Bi$_{35.6}$, Pd$_{5.5}$Pb$_{56.7}$Bi$_{37.8}$, Rh$_{2}$Pb$_{58.8}$Bi$_{39.2}$, Ni$_{11}$Pb$_{53.4}$Bi$_{35.6}$.
All elements
were used as received: Pb shot (99.9\%, Sigma Aldrich), Bi pieces
(99.999\%, Sigma Aldrich), Ni powder (99.9\%, Thermo Scientific), Pd
powder (99.95\%, Thermo Scientific), Rh pieces (99.9\%, Kurt J. Lesker),
and Au powder (99.96+\%, Thermo Scientific). The elements were weighed,
mixed, and loaded into an alumina crucible, which was capped with a
frit-disc and an inverted crucible. The Canfield set was sealed in a
quartz tube that had been purged five times with argon and evacuated to
below 50~mTorr, and the ampules were heated in programmable box furnaces.
For Au and Pd, the charge was melted at 450~$^\circ$C, held for 24~h,
cooled to 200~$^\circ$C at 5~$^\circ$C/h and then to 150~$^\circ$C at
2~$^\circ$C/h, and held for three days before the still-hot ampule was
inverted into a centrifuge and spun. For Rh, the mixture was heated to
1050~$^\circ$C, soaked for two days, quickly cooled to 450~$^\circ$C,
cooled to 200~$^\circ$C at 2~$^\circ$C/h and then to 180~$^\circ$C at
1~$^\circ$C/h, and spun. For Ni, the charge was heated to 500~$^\circ$C,
held for 24~h, cooled to 350~$^\circ$C at 1~$^\circ$C/h, held for three
days, and decanted. Before measurement, residual flux was removed by
polishing and by etching in a mixture of glacial acetic acid and hydrogen
peroxide.

\paragraph{Structural and chemical characterization.}
The crystals were imaged using a scanning electron microscope (SEM) and initial elemental analysis was performed by energy-dispersive X-ray spectroscopy
(EDS) on an FEI Quanta 200 FE-ESEM equipped with an Oxford Instruments
X-Max EDX detector, and the data were processed with the Oxford
Instruments AZtec software.

For precise determination, polished crystals were digested in a 9:1 mixture of trace-metal grade nitric acid (Fisher Chemical) and hydrochloric acid (Fisher Chemical) and diluted with Milli-Q water. Calibration curves of the measured elements were obtained by measuring a dilution series of a mixture of 1 g/L standards of gold, palladium, lead, bismuth and rhodium (Sigma Aldrich).
The ratio of the elements in the resulting solutions was determined using an Agilent 5800 Inductively Coupled Plasma - Optical Emission Spectroscopy (ICP-OES) instrument.
Single-crystal X-ray diffraction (SCXRD) was
carried out at room temperature on either a Rigaku XtaLAB Synergy-S
diffractometer (Mo~$K_\alpha$ source, HyPix-Arc 150 hybrid-pixel
detector) or a Bruker-AXS D8 Venture four-circle diffractometer
(Mo~$K_\alpha$ radiation, APEX2 CCD detector). Frame integration used
CrysAlisPro or APEX2, respectively, and structures were solved and
refined in Jana2020.\cite{petvrivcek2023jana2020} SCXRD was also used to
orient the crystals for the anisotropic measurements described below.

\paragraph{Magnetization.}
Magnetic susceptibility was measured by vibrating-sample magnetometry in
a Quantum Design MPMS3 SQUID magnetometer. Crystals were mounted on
quartz paddles with GE varnish, with the crystallographic $c$ axis either
parallel or perpendicular to the applied field. The dimensions $a,~b,~c$ of the cuboid crystals were measured digitally using a microscope for the calculation of the volume and the demagnetizing factors. The latter were estimated from the rectangular-cuboid
expression $4ab/[4ab+3c(a+b)]$.\cite{prozorov2018effective} To minimize
the effect of remnant fields in the superconducting magnet, each
measurement was preceded by oscillating the magnet to zero field from
2~T at 15~K, well above the $T_c$ of the samples.

\paragraph{Electrical transport.}
Resistivity was measured in a Quantum Design 9~T PPMS using the ETO option. Crystals were fixed
to sapphire substrates with GE varnish and contacted in a four-probe
geometry along the $c$ axis using four gold wires and silver paint. The
mounted samples were placed on either a horizontal rotator or an electrical transport option puck, and the gold wires were soldered to the contact pads. A 1~mA drive
current was applied at frequencies between 3 and 11~Hz.

\paragraph{Heat capacity.}
Heat capacity was measured with the heat-capacity option of a Quantum
and in a high field between 1.8 and
300~K, after which polished crystals of known mass were placed on the
pre-measured grease.

\paragraph{DFT}

Density Functional Theory (DFT) calculations were performed on the experimental crystal structures with Vienna ab initio simulation package (VASP) 6.4.2 \cite{Kresse1993AbMetals, Kresse1994AbGermanium, Kresse1996EfficiencySetb, Kresse1996EfficientSet}. The Perdew--Burke--Ernzerhof (PBE) generalized gradient approximation (GGA)\cite{Perdew1996GeneralizedSimplec} functional was used for exchange-correlation with the recommended potpaw-PBE.64 projector-augmented wave (PAW) pseudopotentials provided by VASP\cite{Blochl1994ProjectorMethod, Kresse_ultrasoft_1999}. The calculations used a converged (criterion of $<10^{-6}$ eV) plane-wave energy cut-off of 550 eV $\Gamma$-centered $11\times11\times11$ Monkhorst-Pack $\mathbf{k}$-point grid \cite{Monkhorst_kpoints_1976} and spin-orbit coupling. Band structures and densities of states were generated with the \texttt{sumo} package.\cite{sumo}

% =====================================================================
\section{Results and Discussion}
% =====================================================================

\subsection{Synthesis, phase formation, and structure}
Single crystals of $M$Pb$_{4-x}$Bi$_x$ ($M$~=~Au, Pd, Rh) adopting the
PtPb$_3$Bi structure type were obtained from a Pb-Bi self-flux. The Pb-Bi
system is well suited for this purpose because it forms a deep eutectic
near 126~$^\circ$C.\cite{gokcen1992bi} Since most solid-state syntheses
are carried out at much higher temperatures, this unusually low-melting
flux, perhaps together with the cost of the late transition metals involved,
may be one reason these phases were previously overlooked.

The successful formation of each compound was first confirmed by SEM-EDS imaging of as-synthesized crystals (Figures~S1-S3), where the $M$Pb$_{4-x}$Bi$_x$ crystals were observed, covered with drops of remnant flux. In order to obtain more precise elemental ratios, ICP-OES measurements were performed on digested solutions of polished crystals. The resulting empirical formulas are AuPb$_{2.78}$Bi$_{1.35}$, PdPb$_{2.08}$Bi$_{1.92}$, RhPb$_{1.56}$Bi$_{2.68}$. The formulas clearly deviate from the perfect $M$Pb$_3$Bi formula. Interestingly, based on the empirical ratios, the total number of valence electrons per formula unit is 20.6, 19.9 and 20.2 for $M$~=~Au, Pd and Rh, or, approximately 20 for all of the compounds, with AuPb$_{4-x}$Bi$_{x}$ being an outlier.
The framework appears to keep the electron count per formula unit nearly constant: a change in the valence-electron count of M is offset by a compensating change in the Pb:Bi ratio (e.g. Pd~$\rightarrow$~Au is balanced by Bi~$\rightarrow$~Pb).
Notably, the number of electrons per formula unit expected in stoichiometric PtPb$_3$Bi is 19. While early reports lack experimental elemental analysis\cite{matkovic1978kristallstruktur}, the EDS results in a recent report on single-crystalline PtPb$_3$Bi\cite{lesser2025learning} suggest an empirical stoichiometry that would result in an electron count between $\approx19.7$ (single point on the crystal) and $\approx20.6$ (for the entire crystal), thus approximately 20 as well. As a result, the $M$Pb$_{4-x}$Bi$_x$ phases are expected to have a homogeneity range, rather than just correspond to a line in the phase diagram, as reported for the PtPb$_3$Bi structure.\cite{matkovic1978kristallstruktur} This suggests that a range of stoichiometries is possible for any given $M$, while conserving the total electron count of 20, that would correspond to a different density of $M$ vacancies and a corresponding Pb/Bi ratio. Finally, the larger deviation of the total electron count in AuPb$_{4-x}$Bi$_{x}$ could either be an experimental artifact, or potentially a manifestation of the aurophilic effect. In the refined crystal structures, the Au-Au dimer distance is 2.79~\AA, well within the aurophilic range.\cite{schmidbaur2008briefing} It is thus potentially capable of hosting excess electron density, similar to previous reports of stabilization of 19-electron ternary phases.\cite{seibel2015gold}

We used SCXRD to verify and solve the crystal structure. The
combination of pronounced Pb-Bi mixing, evident from the elemental analysis, with a lattice made
up almost exclusively of heavy atoms, made the refinements challenging and
sensitive to absorption-correction errors. Nevertheless, we obtained
solutions with $R(I>3\sigma)<5\%$ for all three $M$Pb$_{4-x}$Bi$_x$
structures (Table~S1). Because Pb and Bi have nearly identical scattering
factors, we could not cleanly resolve their site occupancies with a
laboratory X-ray source. 
Accordingly, every Pb/Bi site was modeled as a mixed Pb/Bi position with the relative occupancy fixed to the ICP-derived ratio; refining the M-site occupancy gave no improvement, so it was held fully occupied.
The full crystallographic details are presented in Tables~S1-S3, and the corresponding
CIF files are provided as Supporting Information.

The substitution survey is summarized in Figure~\ref{fig:fig1}b. Together
with the parent Pt, the transition metals Au, Pd, and Rh stabilize the
PtPb$_3$Bi structure type, indicating that the framework tolerates a range
of $M$ radii and $d$-electron counts and accommodates changes in the latter
by adjusting the Pb:Bi ratio. Under the conditions we explored, several
other metals did not form the structure: $M$~=~Ag, Cd, and Ir yielded mainly
Pb$_{0.7}$Bi$_{0.3}$, $M$~=~Cu and Zn yielded mostly crystals of the
elemental metal, and Co, which is highly immiscible in both
Pb\cite{Okamoto-Pb} and Bi,\cite{ishida1990bi} remained largely unreacted up
to 1200~$^\circ$C. We cannot rule out that these or other ternaries could be
stabilized under different synthetic conditions.

The unsuccessful substitutions can be rationalized along chemical lines. For
$M$~=~Ag, Cd, Cu, and Zn, similar to the case of Co, the metals forms no stable binaries with
either Pb or Bi, and no known Pb-Bi ternaries, at ambient
pressure.\cite{LukasAg-Bi-Pb,moser1988bi,liu2009thermodynamic,manasijevic2010prediction,seith1955kenntnis}
The failure of the $3d$ metals most likely reflects an $M$ radius
that is too small, and, in the case of Ag, a covalent radius that is too large. The
failure of Cd may instead be a manifestation of the 20-electron conservation rule, since
no positive $x$ satisfies the rule for CdPb$_{4-x}$Bi$_x$. Ir is the most
surprising absence: from the perspective of chemical intuition, and given the existence of stable
Ir binaries with both Pb and Bi, one would expect it to behave like the
metals that do form the PtPb$_3$Bi structure. Such a compound may well exist,
but its synthesis could be kinetically hindered by the low solubility of Ir
in both Pb and Bi; indeed, even preparing new binary Ir phases can require
multistep routes and spectator species.\cite{isaeva2015structure}

A distinct structural outcome occurs for the smallest metal we tested.
When $M$~=~Ni, the product is a Pb-substituted version of the known
superconductor NiBi$_3$, namely NiBi$_{3-x}$Pb$_x$ (Figure~S4), rather
than the $M$Pb$_3$Bi type.\cite{silva2013superconductivity} The two structures are
nonetheless related: as the $M$ radius decreases from Pd to Ni, the metal
can no longer support the octagonal cage by forming planar dimers, and
the cage collapses into a hexagonal channel in which the Ni atoms form
zigzag chains along the $c$ axis. For the refinement of the crystal structure of NiBi$_{3-x}$Pb$_x$ (Table~S1), we again fixed the Bi/Pb ratio to the value obtained from EDS (Figure S4). From magnetic susceptibility
(Figure~S7) we find $T_c\approx2.2$~K for NiBi$_{3-x}$Pb$_x$, lower than the reported $T_c\approx4$~K.

Under high pressure conditions, Co has been reported to form the NiBi$_3$-like CoBi$_3$ structure, which also superconducts with a reported $T_c\approx0.5$~K.\cite{schwarz2013cobi, tence2014cobi3} In addition, other high-pressure binaries of Bi with both Co\cite{badding2025predicted} and Cu\cite{clarke2017creating} are known to exist. The resulting structures are usually described as chains of the transition metal segregated from pillars of the Bi atoms, reminiscent of the complete segregation of the elements at ambient conditions. As some of the aforementioned high-pressure phases are metastable under ambient pressure, and could be isolated and characterized, high-pressure synthesis may therefore provide a route to additional $M$Pb$_{4-x}$Bi$_x$ compounds. Taken together with the aforementioned NiBi$_3$-like RhBi$_3$, as well as the isotypic La$_2$NiBi\cite{ott2019structure} and La$_2$NiSb\cite{schaefer2014la2nisb}, we believe that the boundaries of $M$-Pb-Bi and related phases can be expanded even further.

\subsection{Superconducting properties}
Representative longitudinal resistivities between room temperature and
1.8~K are shown in Figure~\ref{fig:fig1}c. All three compounds are poor
metals, with residual resistivity ratios $RRR\approx1.3$ to $1.5$, and
each shows a sharp drop to zero resistivity between 3 and 5~K. No
additional transition from residual Pb or Pb-Bi alloy is observed, which,
together with the etching procedure, indicates that the signal originates
in the bulk crystals.

\begin{figure*}[ht]
\includegraphics{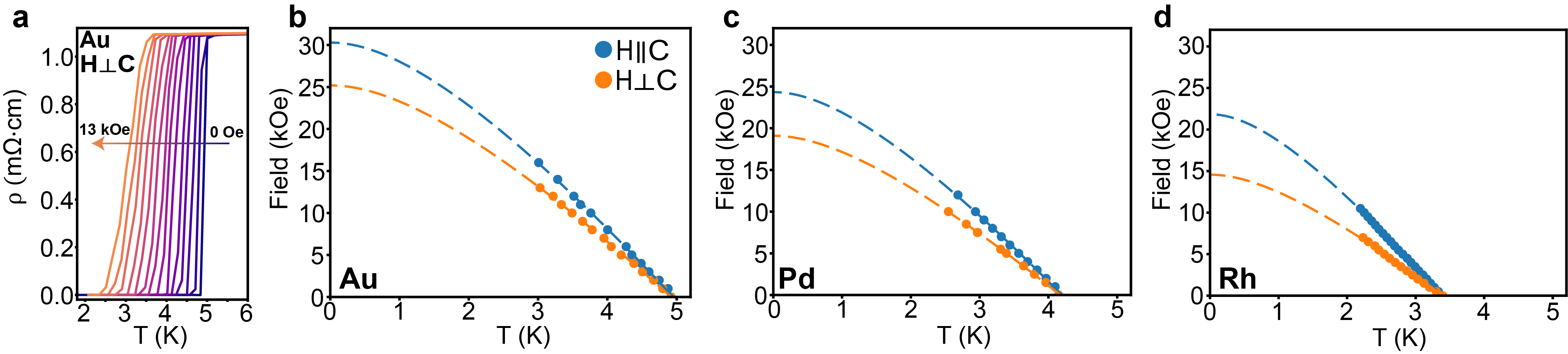}
\caption{\label{fig:fig2} (a) Temperature-dependent resistivity of
AuPb$_{4-x}$Bi$_x$ under increasing magnetic field at 1~kOe increments perpendicular to the
$c$ axis, shown as a representative example. (b to d) Upper critical field
phase diagrams obtained by fitting the experimental points (dots) to the
WHH model (dashed lines) for $M$~=~Au, Pd, and Rh, for fields parallel
(blue) and perpendicular (orange) to the $c$ axis.}
\end{figure*}

The tetragonal, channel-based structure naturally led us to examine the
anisotropy of the superconducting state. Figure~\ref{fig:fig2} shows the
upper critical field phase diagrams, constructed by tracking the
temperature of a 50\% drop in resistivity from the normal-state value as
a function of field (Figure~\ref{fig:fig2}a), which defines
$H_{c2}(T)$, for the two principal field orientations; the raw curves for
all orientations are given in Figure~S5. The Au compound shows the
highest $T_c$ and the largest critical fields (Figure~\ref{fig:fig2}b),
followed by Pd (Figure~\ref{fig:fig2}c) and Rh
(Figure~\ref{fig:fig2}d). For all three, $H_{c2}$ lies well below the
weak-coupling Pauli limit, $\mu_0 H_p\approx1.84\,T_c$ (about 90~kOe for
$T_c\approx5$~K).\cite{clogston1962upper} This places the materials in
the orbitally limited regime, so that spin-paramagnetic pair breaking can
be neglected despite the strong spin-orbit coupling expected from the
heavy constituents. The low $RRR$ furthermore places the samples in the
dirty limit. We therefore fit the data to the orbitally limited
Werthamer-Helfand-Hohenberg (WHH)
equation,\cite{helfand1966temperature} without spin or spin-orbit terms,
\[
\ln\!\left(\frac{1}{t}\right)=
\psi\!\left(\frac{1}{2}+\frac{\bar{h}}{2t}\right)-\psi\!\left(\frac{1}{2}\right),
\]
where $\psi$ is the digamma function, $t=T/T_{c}$, and
$\bar{h}=4H_{c2}/[\pi^{2}T_{c}\,|\mathrm{d}H_{c2}/\mathrm{d}T|_{T_{c}}]$.

Extrapolating to zero temperature gives $H_{c2}^{\parallel c}(0)=30$,
$24$, and $22$~kOe for Au, Pd, and Rh, respectively. Within the anisotropic
Ginzburg-Landau (GL) theory,\cite{clem1998anisotropy} the corresponding
$H_{c2}^{\perp c}$ values yield modest anisotropy ratios
$\gamma_{\xi}(0)=H_{c2}^{\parallel c}(0)/H_{c2}^{\perp c}(0)=1.2$, $1.3$,
and $1.5$ for Au, Pd, and Rh. A longer coherence length along $c$ is
consistent with the quasi-one-dimensional character of the structure. The
GL coherence lengths follow from
$H_{c2}^{\parallel c}=\Phi_{0}/(2\pi\xi_{ab}^{2})$ and
$H_{c2}^{\perp c}=\Phi_{0}/(2\pi\xi_{ab}\xi_{c})$, with $\Phi_0$ the flux
quantum, and are collected in Table~\ref{tab:SCparams} along with the
other quantities discussed below.

\begin{figure*}[t]
\includegraphics{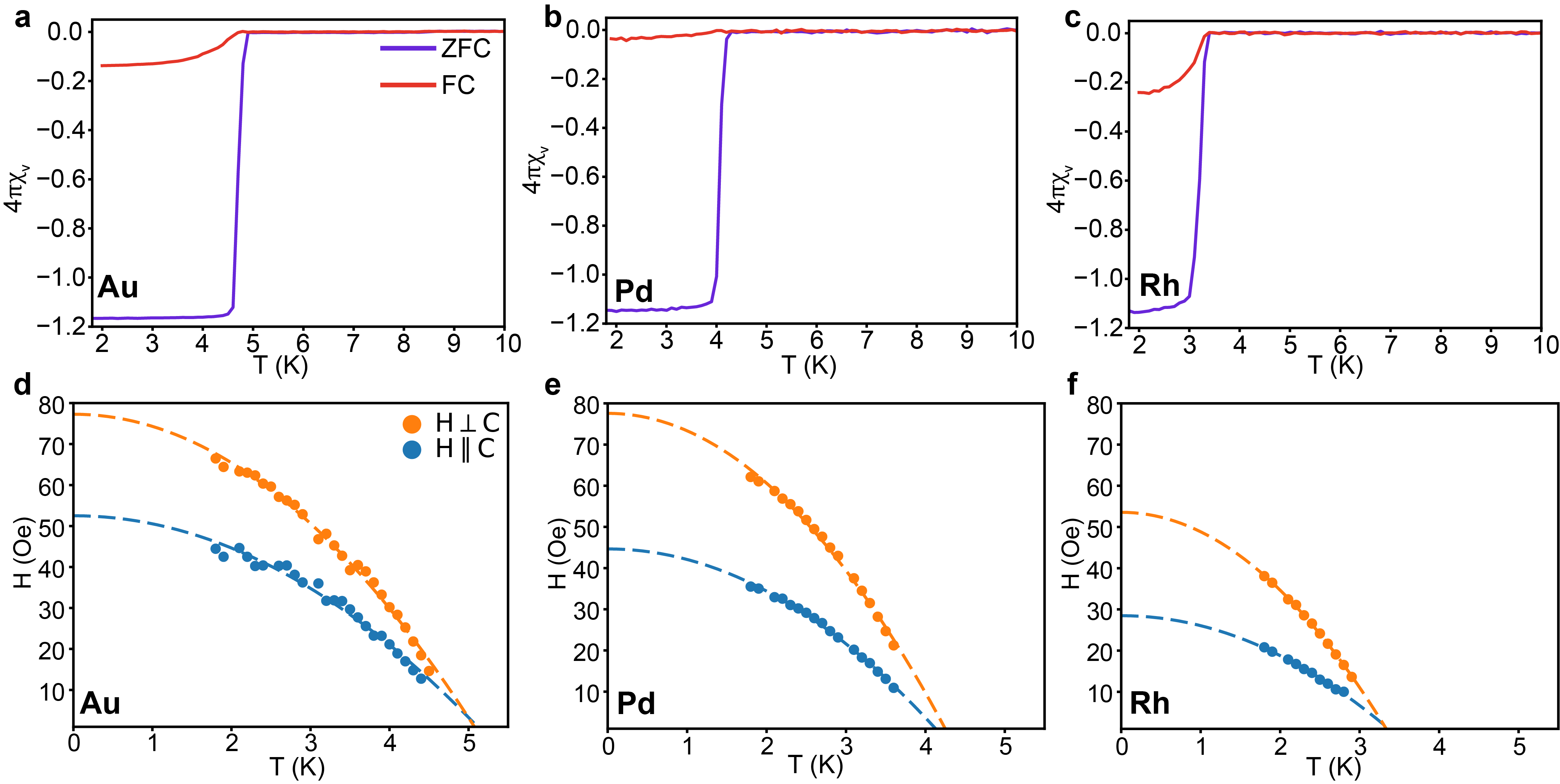}
\caption{\label{fig:fig3} (a-c) Magnetic susceptibility under a 5~Oe
bias field for $M$~=~Au, Pd, and Rh under zero-field-cooled (blue) and
field-cooled (orange) conditions. (d-f) Lower critical field phase
diagrams, with $H_{c1}(T)$ extracted from the Meissner curves (dots) and
fit to $H_{c1}(T)=H_{c1}(0)[1-(T/T_c)^2]$ (dashed lines) for $H\perp c$
(red) and $H\parallel c$ (purple).}
\end{figure*}

To confirm that the superconductivity is a bulk property, we measured the
magnetization across $T_c$. Demagnetization-corrected zero-field-cooled
(ZFC) and field-cooled (FC) susceptibilities under a 5~Oe field parallel
to the $c$ axis (Figure~\ref{fig:fig3}a to c) show strong diamagnetic
shielding, with the ZFC signal approaching full flux expulsion, and no
additional transition from Pb or Pb-Bi alloys. The ZFC demagnetization-corrected shielding fractions are found to be $4\pi\chi_v\approx$~-1.1~to~-1.2, slightly exceeding the ideal full-shielding value. This modest over-correction is attributed to uncertainty in the sample dimensions and the cuboidal approximation\cite{prozorov2018effective} used for the demagnetization factor, and is consistent with full, bulk shielding within geometric uncertainty.

We then determined the lower critical field $H_{c1}$ from the deviation
of the Meissner curves from linearity at several temperatures and both
orientations (Figure~S6). The resulting phase diagrams
(Figure~\ref{fig:fig3}d-f) are fit to
$H_{c1}(T)=H_{c1}(0)[1-(T/T_c)^2]$ to extract $H_{c1}(0)$. The penetration
depth along principal axis $i$, $\lambda_i$, then follows from the
tetragonal GL relations\cite{clem1989phenomenological}
$H_{c1}^{\parallel c}=\frac{\Phi_{0}}{4\pi\lambda_{ab}^{2}}(\ln\kappa_{c}+0.497)$
and
$H_{c1}^{\perp c}=\frac{\Phi_{0}}{4\pi\lambda_{ab}\lambda_{c}}(\ln\kappa_{ab}+0.497)$,
with $\kappa_{c}=\lambda_{ab}/\xi_{ab}$ and
$\kappa_{ab}=\sqrt{\lambda_{ab}\lambda_{c}/(\xi_{ab}\xi_{c})}$. As
expected from anisotropic GL theory, the penetration-depth anisotropy is
the inverse of the coherence-length anisotropy,
$\gamma_{\lambda}=\gamma_{\xi}^{-1}$, and Table~\ref{tab:SCparams} shows
that the measured $\gamma_{\lambda}^{-1}$ values follow this trend.
Finally, all of the GL parameters satisfy $\kappa\gg1/\sqrt{2}$, confirming
that the compounds are type-II superconductors.

\begin{figure}[h]
\includegraphics{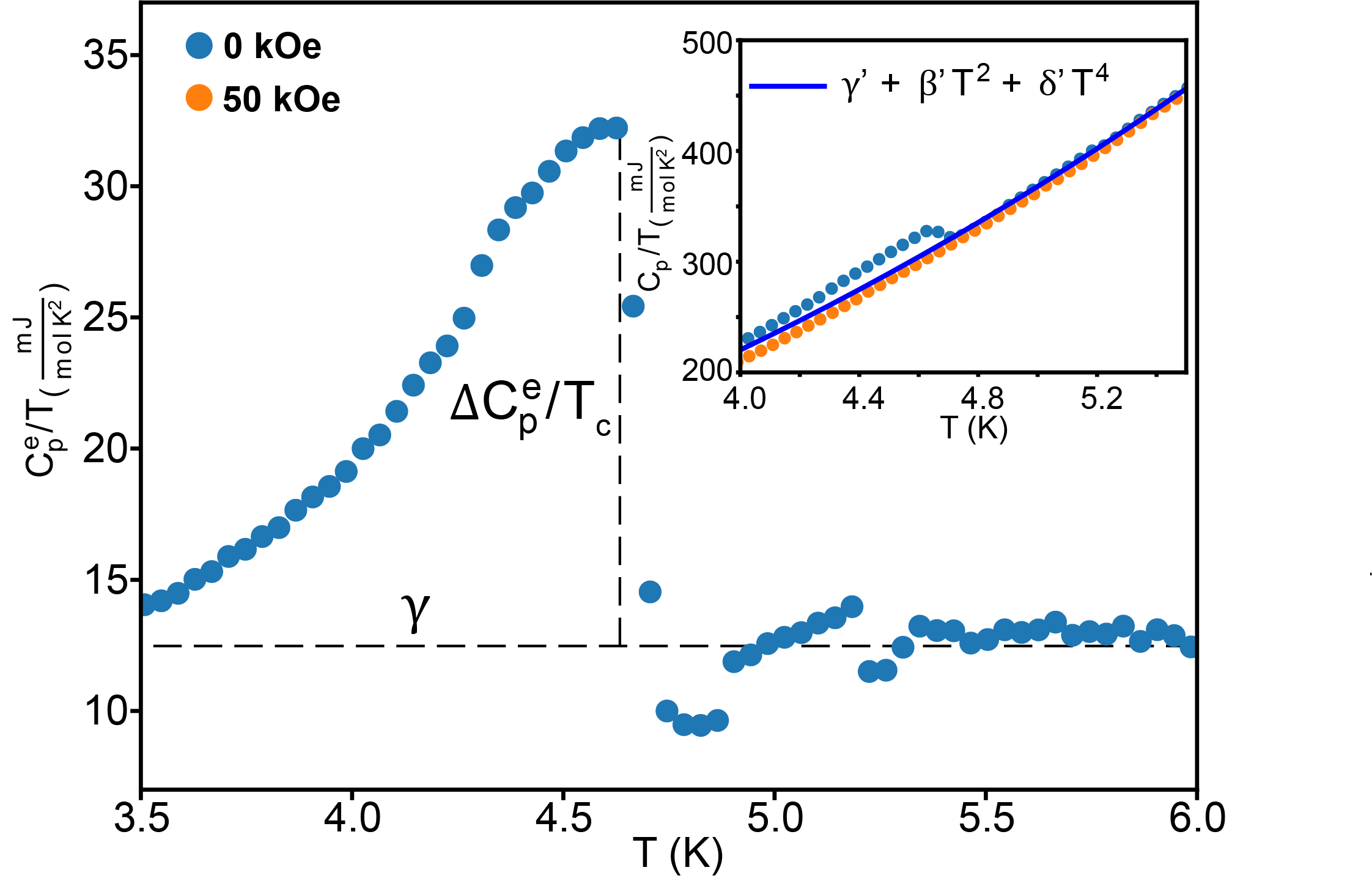}
\caption{\label{fig:fig4} Electronic specific heat $C_p^e/T$ of
AuPb$_{4-x}$Bi$_x$ across the superconducting transition. Dashed lines
mark the Sommerfeld coefficient $\gamma$, extracted from 50~kOe data at low
temperature, and the jump $\Delta C_p^e/T_c$. The inset shows the raw total
specific heat $C_p/T$ in zero and high field, with the high-temperature fit as the blue line.}
\end{figure}

\begin{table*}[!htbp]
\caption{\label{tab:SCparams}
Superconducting parameters of the $M$Pb$_{4-x}$Bi$_x$ compounds for
fields parallel and perpendicular to the $c$ axis. For ease of comparison
with $\gamma_\xi$, $\gamma_\lambda^{-1}$ is tabulated. For $\xi$,
$\lambda$, and $\kappa$, $\parallel c$ and $\perp c$ refer to the $c$ and
$ab$ axes, respectively.}
\setlength{\tabcolsep}{4pt}
\resizebox{\textwidth}{!}{%
\begin{tabular}{llcccccccc}
\toprule
Element & Orientation
& $T_c$ (K)
& $H_{c2}$ (kOe)
& $\xi$ (nm)
& $\gamma_{\xi}$
& $H_{c1}$ (Oe)
& $\lambda$ (nm)
& $\gamma_{\lambda}^{-1}$
& $\kappa$ \\ \midrule
\multirow{2}{*}{Au}
& $\parallel c$ & \multirow{2}{*}{4.9} & 30 & 12.5 & \multirow{2}{*}{1.2} & 53 & 222 & \multirow{2}{*}{1.6} & 34 \\
& $\perp c$     &                      & 25 & 10.0 &                      & 77 & 355 &                      & 24 \\
\multirow{2}{*}{Pd}
& $\parallel c$ & \multirow{2}{*}{4.2} & 24 & 14.8 & \multirow{2}{*}{1.3} & 45 & 196 & \multirow{2}{*}{1.9} & 33 \\
& $\perp c$     &                      & 19 & 11.6 &                      & 78 & 383 &                      & 21 \\
\multirow{2}{*}{Rh}
& $\parallel c$ & \multirow{2}{*}{3.4} & 22 & 18.4 & \multirow{2}{*}{1.5} & 28 & 224 & \multirow{2}{*}{2.2} & 40 \\
& $\perp c$     &                      & 15 & 12.3 &                      & 54 & 491 &                      & 22 \\
\bottomrule
\end{tabular}%
}
\end{table*}

As a final and independent test of bulk superconductivity, we measured
the specific heat across $T_c$ in zero field and at 50~kOe, above the upper critical fields extrapolated for the compounds. The
Bardeen-Cooper-Schrieffer (BCS) theory predicts a jump in the electronic
specific heat of $\Delta C^e_p/(\gamma T_c)\approx1.43$, where $\gamma$ is
the Sommerfeld coefficient. As fitting the zero-field, high-temperature data
to $C_p/T=\gamma+\beta T^2+\delta T^4$ consistently returned unphysical
parameters, we used a previously reported method for the related NiBi$_3$,\cite{fujimori2000superconducting} estimating $\gamma$ from 50~kOe measurements (well above
$H_{c2}$) between 1.8 and 2.3~K (Figure~S8). 

We obtained $\gamma$ values of 12.8, 12.3, and
8.6~mJ\,mol$^{-1}$\,K$^{-2}$ for Au, Pd, and Rh, values typical of
intermetallics, and generally tracking with the observed $T_c$.
By fitting zero-field data just above the transition temperature to the same equation and subtracting it from the data points at the transition we could get the change in the heat capacity $\Delta C_p$. By using the $\gamma$ values previously obtained from the 50~kOe, low temperature data, we find $\Delta C_p/(\gamma T_c)=1.50$, $1.58$, and
$1.60$ for Au, Pd, and Rh. These values, near or slightly above the BCS
expectation, indicate bulk superconductivity with moderate coupling
strength. Figure~\ref{fig:fig4} illustrates the analysis for Au, where the
high-field, low-temperature $\gamma$ defines the baseline from which the
jump $\Delta C$ is measured; the inset shows the high-temperature Debye
fit and the full suppression of the transition by a 50~kOe field. The
corresponding analyses for Pd and Rh appear in Figures~S9 and S10. Note the uncertainty in the specifics of the heat capacity analysis due to artifacts seen in some samples (shown in Figures~S9-S10). Nonetheless the heat capacity confirms bulk superconductivity. Taken
together with the shielding fractions from magnetization, the specific
heat establishes the bulk nature of the superconducting state.

\begin{figure*}[h!]
\includegraphics{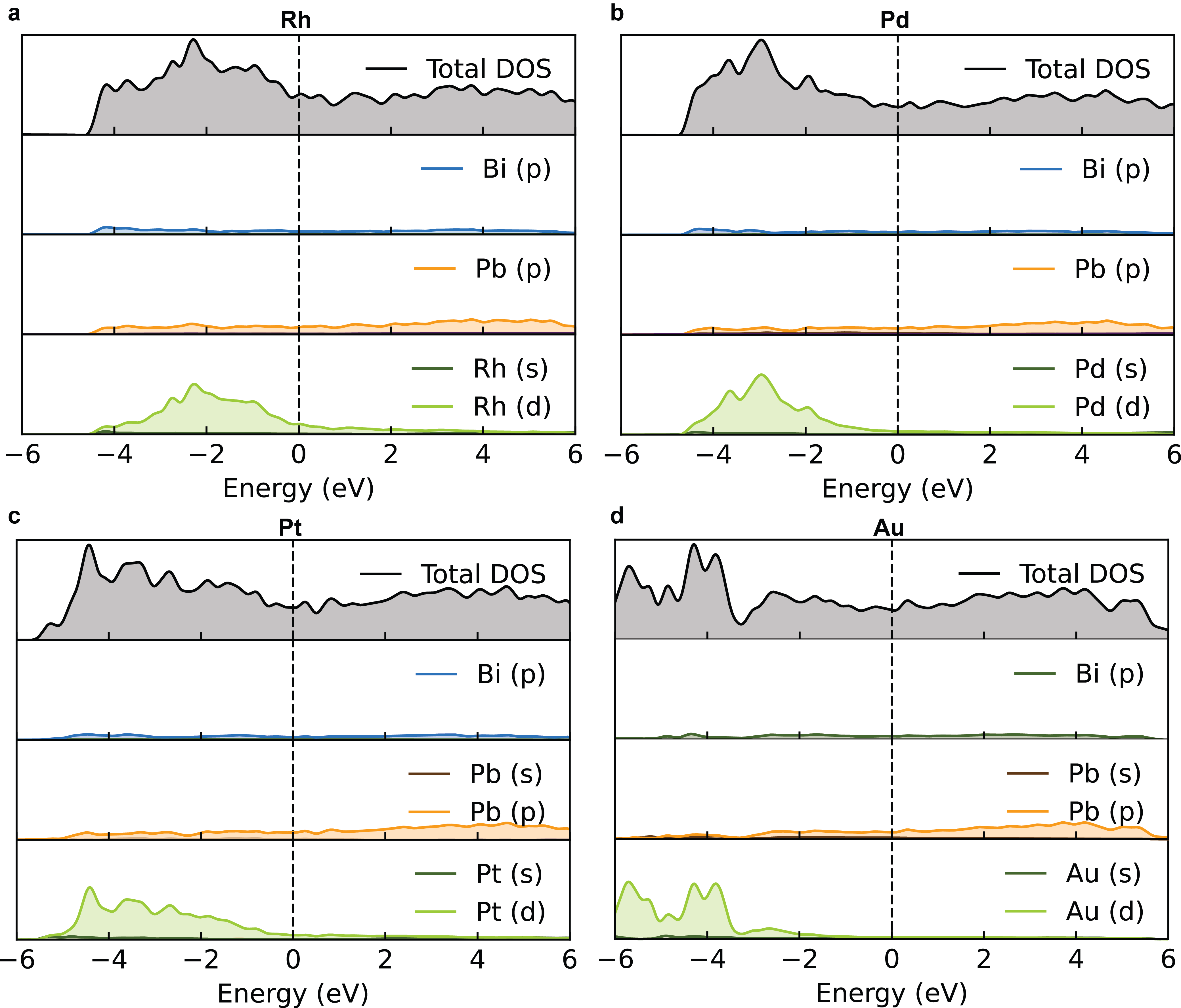}
\caption{\label{fig:fig5} Calculated total (black lines) and partial (colored lines) density of states for $M$Pb$_3$Bi structures with an idealized stoichiometry for $M$~=~Rh, Pd, Pt, Au (a-d, respectively).}

\end{figure*}

\subsection{DFT Calculations.}
DFT calculations were performed in order to gain further insight into the structural and electronic properties of the compounds, as well as of PtPb$_3$Bi (from ICSD\cite{ICSD} 58834\cite{matkovic1978kristallstruktur}). Since the actual Pb-Bi distribution is not known, for the sake of comparison, we assumed perfect $M$Pb$_3$Bi structures, where the mixed Pb/Bi were replaced with a fully occupied site of either Pb or Bi to mimic a stoichiometrically perfect PtPb$_3$Bi. In addition, we used the experimental structures as solved and refined from the SCXRD measurements. As expected from the dimensionality and the bonding in the compounds, as well as the properties measured, the resulting band structures (Figure S11) show significantly dispersive bands in the trajectories in the crystallographic $c$ axis ($\Gamma\rightarrow \text{Z}$), while remaining relatively flat within the crystallographic $ab$ plane ($\Gamma\rightarrow \text{X} \rightarrow \text{M}$; $\text{Z} \rightarrow \text{R} \rightarrow \text{A}$).

Due to our computational method, the $M$Pb$_3$Bi structures used for the calculation are expected to be either electron-poor or electron-rich with respect to the real structure. We can examine the resulting total and partial densities of states (DOS) of the dominant atomic orbitals, shown in Figure~\ref{fig:fig5}, understand adjusting the Fermi level as analogous to adjusting the experimental Pb/Bi ratio. In reality, the real chemical composition and the location of the Fermi energy usually cannot be cleanly disentangled from the structure, which in turn would change the bands location and dispersion. Still, several observations can be made: in the case of $M$~=~Rh, the Fermi level clearly sits on a local maximum of the total DOS, and can be moved to a global minimum, located at an energy that is $\approx0.7$~eV higher, by addition of electrons - effectively replacing some Pb atoms by Bi. For $M$~=~Pt and Pd, the general profile of the total DOS around the Fermi energy is quite similar, as expected from them being part of the same group. Finally, when $M$~=~Au, the resulting Fermi level is located $\approx2$~eV above the Au bands, suggestive again of a clear over-filling of electrons. In addition, the large total DOS several eV below $E_F$, contributed mainly by Au, is suggestive of the contribution of the aurophilic effect, potentially allowing the stabilization of AuPb$_{4-x}$Bi$_x$ with excess electrons as compared to the suggested 20 electron per formula unit.

% =====================================================================
\section{Conclusions}
% =====================================================================

In summary, we have established $M$Pb$_{4-x}$Bi$_x$ as a new family of
superconductors based on the PtPb$_3$Bi structure type, comprising
AuPb$_{4-x}$Bi$_x$, PdPb$_{4-x}$Bi$_x$, and RhPb$_{4-x}$Bi$_x$, all bulk,
moderately anisotropic type-II superconductors with $T_c$ between 3.4 and
4.9~K. The substitution survey defines an initial stability window in which
Au, Pd, Rh, and Pt form the structure, whereas several other transition
metals do not under the conditions we explored. A unifying thread is the
electron count: across the series the Pb:Bi ratio adjusts to offset the
changing valence-electron count of $M$, holding the total near 20 electrons
per formula unit. This points to a homogeneity range rather than a line
phase, rationalizes the absence of a CdPb$_{4-x}$Bi$_x$ member (for which no
composition reaches 20 electrons), and agrees with our DFT calculations,
which indicate that $M$ acts largely by setting the position of the Fermi
level relative to features in the density of states; the same picture is
mirrored in the specific heat, where the Sommerfeld coefficient tracks $T_c$
across the series.

For the smaller Ni, the octagonal channels collapse into the
Pb-substituted NiBi$_3$ structure, and the existence of high-pressure
$M$-Bi binaries such as CoBi$_3$ suggests that pressure could stabilize
still more members of the broader $M$-Pb-Bi family. Looking ahead, resolving
the Pb and Bi site occupancies by synchrotron or neutron diffraction,
extending the upper critical field measurements to higher fields, and
deepening the electronic-structure and bonding analysis should turn the
empirical, electron-count-based stability window reported here into a
predictive design rule and guide the search for higher-$T_c$ members of the
family.

% =====================================================================
\begin{acknowledgement}
This work was funded by the National Science Foundation through the AI
Research Institutes program (Award No.~DMR-2433348) and through a CAREER
award (DMR-2144295), which provided summer research support for A.M. S.A. acknowledges support by the Weizmann Institute of Science Women's Postdoctoral Career Development Award. G.C. is supported by the NSF Graduate Research Fellowship Program under grant number DGE-2039656. The
authors acknowledge use of the Imaging and Analysis Center (IAC) operated
by the Princeton Materials Institute, supported in part by the Princeton
Center for Complex Materials (PCCM), a National Science Foundation
Materials Research Science and Engineering Center (MRSEC; DMR-2011750). The simulations presented in this article were performed on computational resources managed and supported by Princeton University's Research Computing. L.V. thanks Dr. Julius Oppenheim for fruitful discussions.
\end{acknowledgement}

\begin{suppinfo}
EDS spectra for all compounds (Figures~S1 to S4); crystallographic,
atomic-position, and anisotropic displacement parameter tables
(Tables~S1 to S3); raw field-dependent resistivity curves (Figure~S5);
Meissner curves used to extract $H_{c1}$ (Figure~S6); susceptibility of
NiBi$_{3-x}$Pb$_x$ (Figure~S7); additional specific heat data and fits
(Figures~S8 to S10); and band structures calculated by DFT (Figure~S11).\\
Crystallographic Information Files have been deposited
in the CCDC with deposition numbers CSD 2562541-2562544. CIF files for all four structures are also provided.
\end{suppinfo}

\bibliography{refs}

% =====================================================================
% Supporting Information appended for arXiv
% =====================================================================
\clearpage
\setcounter{page}{1}
\renewcommand{\thepage}{S\arabic{page}}
\setcounter{figure}{0}
\renewcommand{\thefigure}{S\arabic{figure}}
\setcounter{table}{0}
\renewcommand{\thetable}{S\arabic{table}}
\setcounter{equation}{0}
\renewcommand{\theequation}{S\arabic{equation}}

\section*{Supporting Information}

% ---------------------------------------------------------------------
% EDS (Figures S1-S4)
% ---------------------------------------------------------------------
\begin{figure}[h]
\centering
\includegraphics[width=\textwidth]{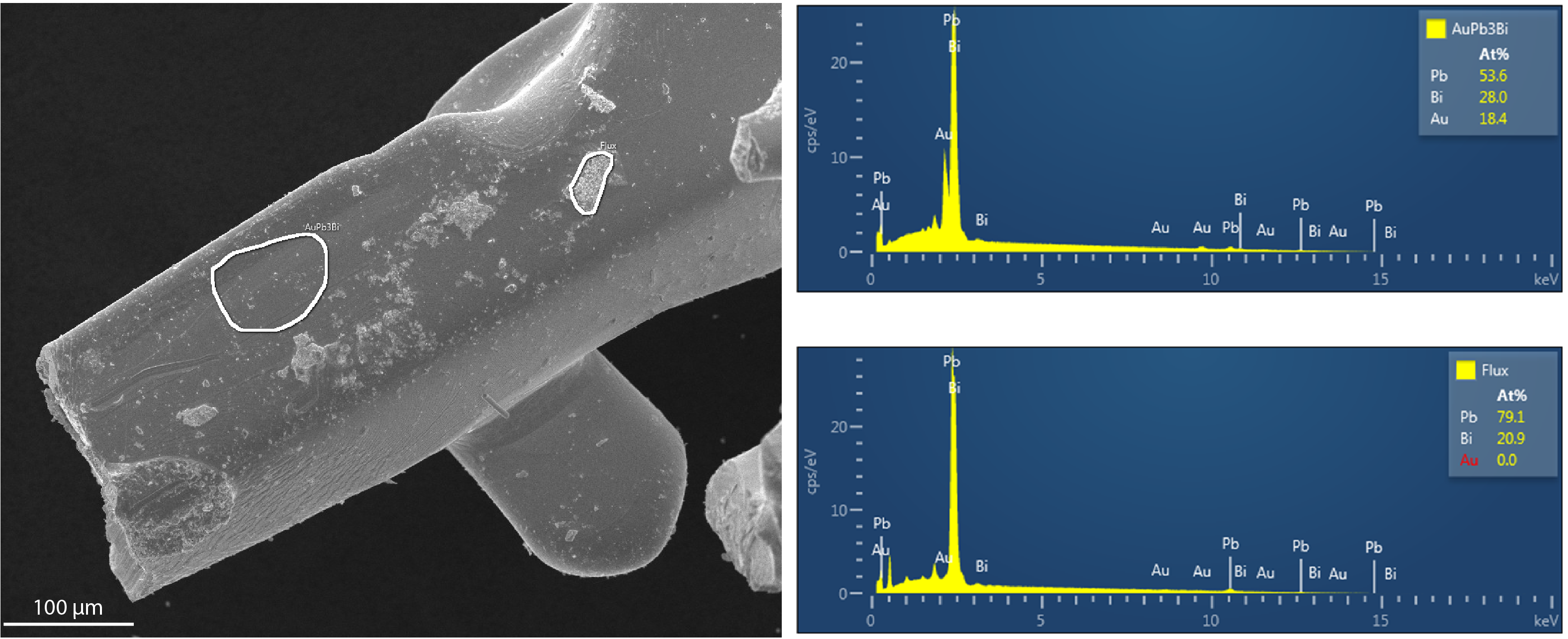}
\caption{EDS spectra of selected clean and flux-covered regions of
AuPb$_{4-x}$Bi$_x$, yielding the formula AuPb$_{2.9}$Bi$_{1.5}$.}
\label{fig:S1}
\end{figure}

\begin{figure}[h]
\centering
\includegraphics[width=\textwidth]{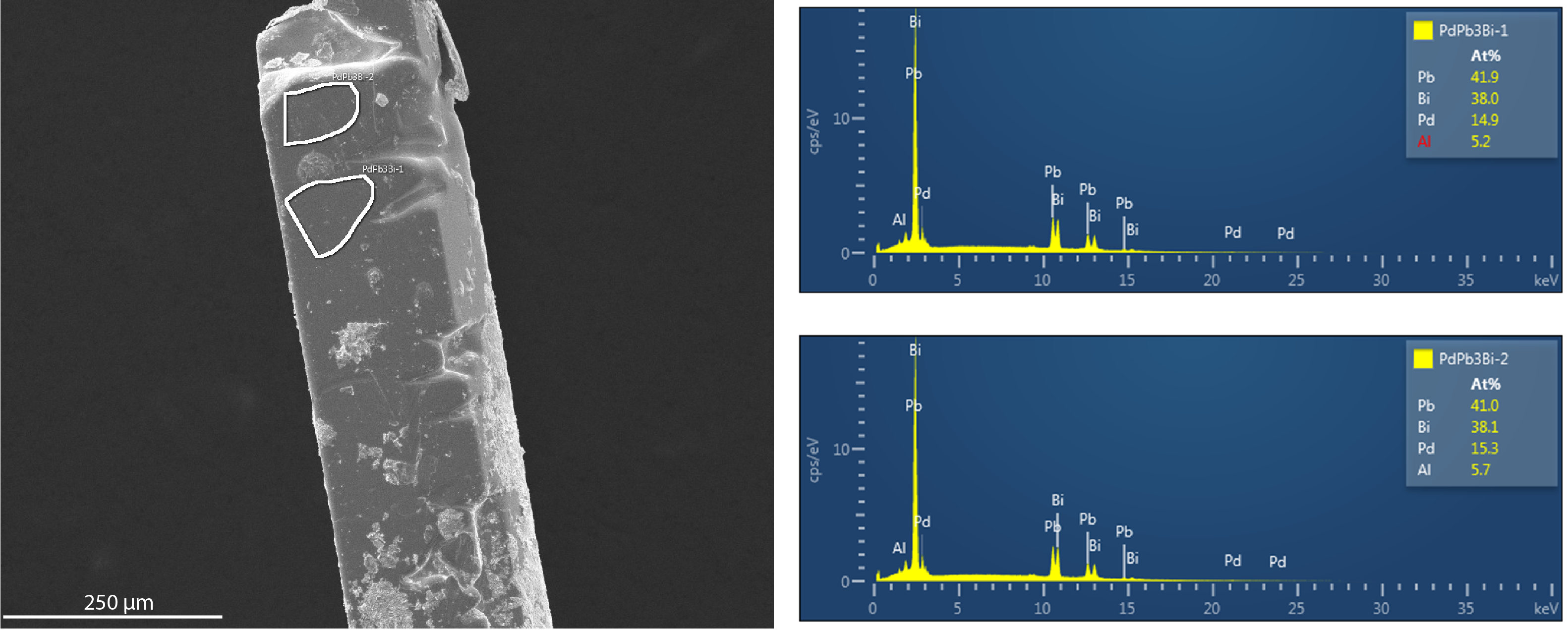}
\caption{EDS spectra of selected clean and flux-covered regions of
PdPb$_{4-x}$Bi$_x$, yielding the formula PdPb$_{2.8}$Bi$_{2.5}$. Counts
from Al in the sample holder are also visible.}
\label{fig:S2}
\end{figure}

\begin{figure}[h]
\centering
\includegraphics[width=\textwidth]{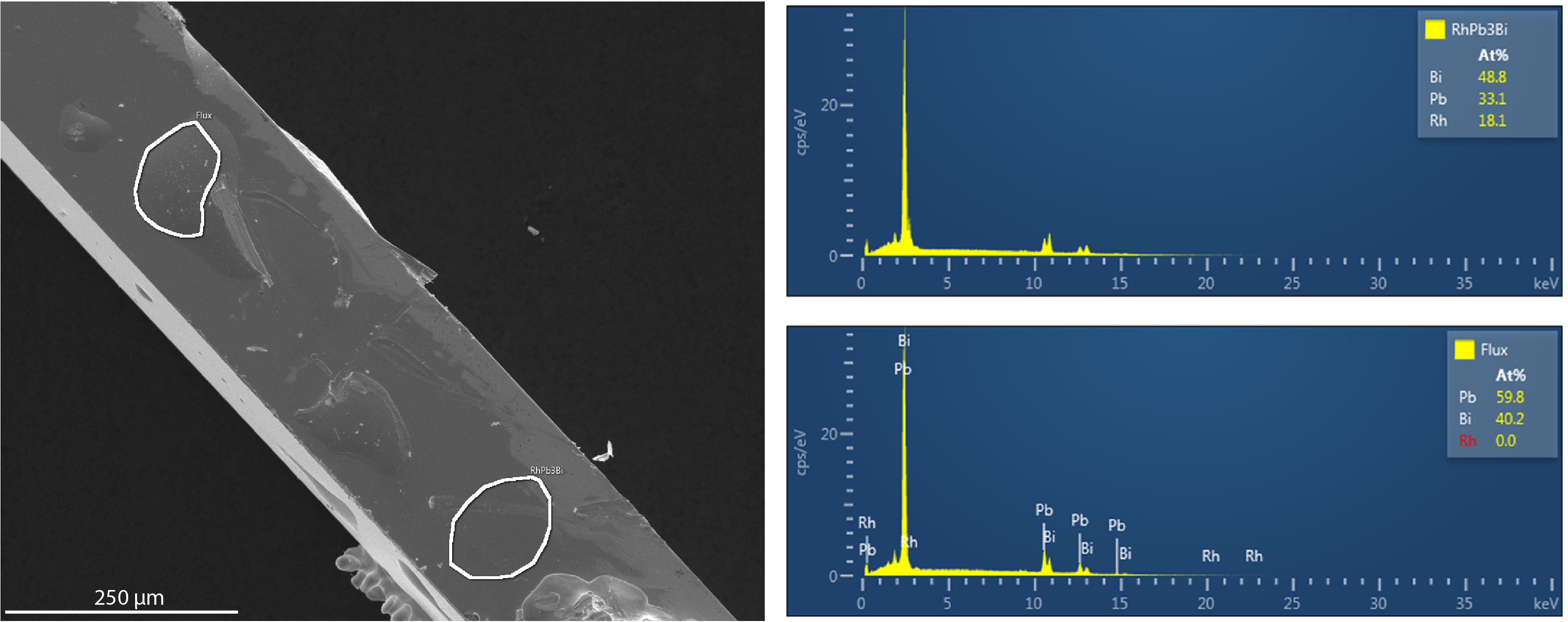}
\caption{EDS spectra of selected clean and flux-covered regions of
RhPb$_{4-x}$Bi$_x$, yielding the formula RhPb$_{1.8}$Bi$_{2.7}$.}
\label{fig:S3}
\end{figure}

\begin{figure}[h]
\centering
\includegraphics[width=\textwidth]{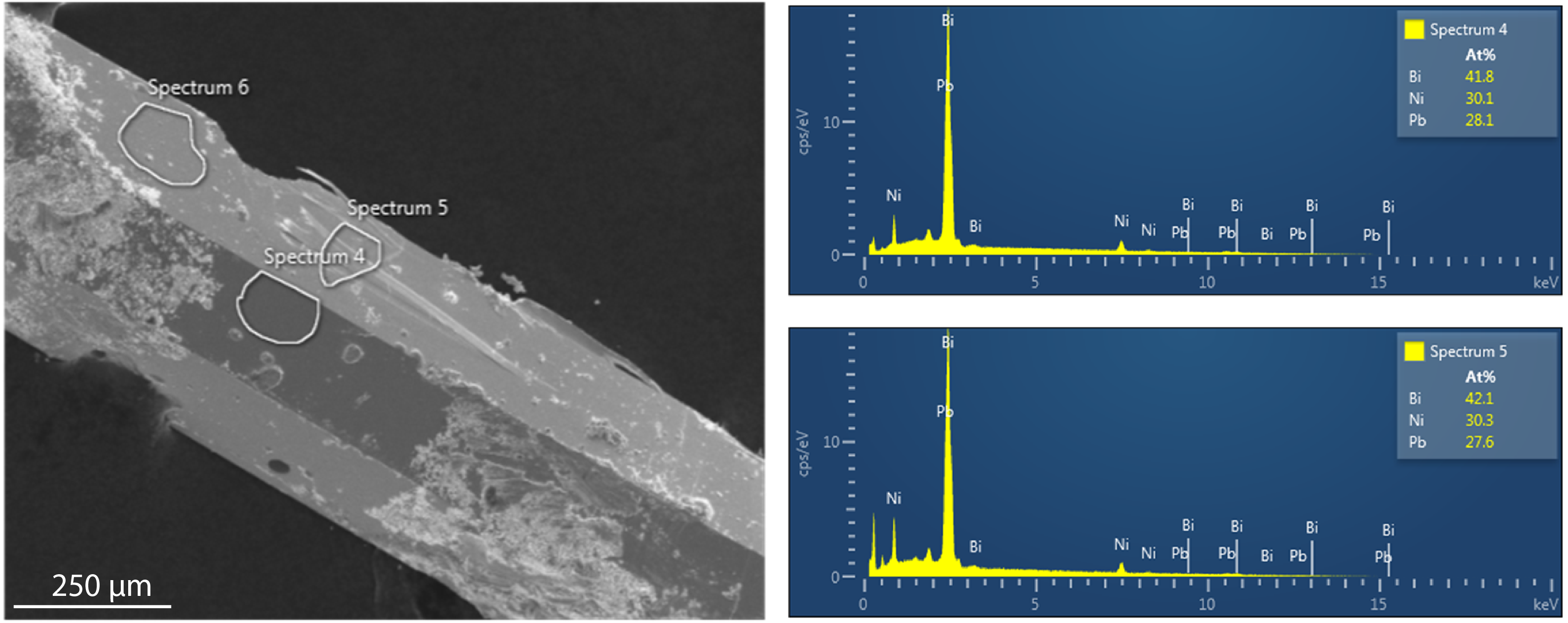}
\caption{EDS spectra of selected clean and flux-covered regions of
NiBi$_{3-x}$Pb$_x$, yielding the formula NiBi$_{1.4}$Pb$_{0.9}$.}
\label{fig:S4}
\end{figure}

% ---------------------------------------------------------------------
% Crystallographic tables (Tables S1-S3)
% ---------------------------------------------------------------------
\begin{table*}[ht]
\centering
\scriptsize
\caption{\label{tab:si_xtalparams}Crystallographic information for the
reported $M$-Pb-Bi structures.}
\begin{tabular}{lcccc}
\toprule
& Pd--Bi--Pb & Rh--Bi--Pb & Au--Bi--Pb & Ni--Bi--Pb \\
\midrule
Crystal family & Tetragonal & Tetragonal & Tetragonal & Orthorhombic \\
CCDC deposit ID
& 2562544
& 2562543
& 2562542
& 2562541\\
Refined composition
& PdPb$_2$Bi$_2$
& RhPb$_{1.48}$Bi$_{2.52}$
& AuPb$_{2.68}$Bi$_{1.32}$
& NiPb$_{1.20}$Bi$_{1.80}$ \\
Crystal dimensions (mm)
& $0.680 \times 0.089 \times 0.031$
& $0.635 \times 0.060 \times 0.048$
& $0.067 \times 0.039 \times 0.027$
& $0.575 \times 0.110 \times 0.067$ \\
Radiation, $\lambda$ (\AA)
& Mo K$\alpha$, 0.71073
& Mo K$\alpha$, 0.71073
& Mo K$\alpha$, 0.71073
& Mo K$\alpha$, 0.71073 \\
Absorption correction
& Gaussian
& Analytical
& Gaussian
& Gaussian \\
Data collection temp. (K)
& 293.15
& 293.15
& 293
& 293.15 \\
Space group
& $P4_2/mnm$
& $P4_2/mnm$
& $P4_2/mnm$
& $Pnma$ \\
$a$ (\AA)
& 11.4966(3)
& 11.57429(16)
& 11.5012
& 8.87104(11) \\
$b$ (\AA)
& 11.4966(3)
& 11.57429(16)
& 11.5012
& 4.04709(6) \\
$c$ (\AA)
& 4.09250(12)
& 3.95324(5)
& 4.297
& 11.47988(12) \\
$\alpha,\beta,\gamma$ ($^\circ$)
& 90, 90, 90
& 90, 90, 90
& 90, 90, 90
& 90, 90, 90 \\
Cell volume (\AA$^3$)
& 540.91(2)
& 529.593(12)
& 568.3968
& 412.149(9) \\
Absorption coefficient (mm$^{-1}$)
& 129.973
& 130.991
& 145.397
& 129.749 \\
$\theta_{\min},\theta_{\max}$ ($^\circ$)
& 2.51, 30.48
& 2.49, 36.27
& 2.50, 37.48
& 2.90, 30.47 \\
Refinement method
& $F^2$
& $F^2$
& $F^2$
& $F^2$ \\
$R_{\mathrm{int}}$ (all, %)
& 19.42
& 14.38
& 11.38
& 12.52 \\
Number of reflections
& 25247
& 47785
& 18063
& 16308 \\
Number of parameters
& 20
& 20
& 20
& 26 \\
Unique reflections $(I>3\sigma(I),\ \mathrm{all})$
& 417, 492
& 681, 756
& 561, 865
& 667, 708 \\
$R(I>3\sigma)$, $wR(I>3\sigma)$ (\%)
& 2.60, 6.19
& 3.38, 9.56
& 3.12, 7.89
& 2.88, 7.82 \\
$R(\mathrm{all})$, $wR(\mathrm{all})$ (\%)
& 3.08, 6.28
& 3.75, 9.71
& 6.40, 9.44
& 3.09, 7.93 \\
$S(I>3\sigma)$, $S(\mathrm{all})$
& 1.8130, 1.6885
& 1.9426, 1.8688
& 1.0410, 0.9998
& 1.8188, 1.7866 \\
$\Delta\rho_{\max},\Delta\rho_{\min}$ (e \AA$^{-3}$)
& 1.37, -1.36
& 1.68, -1.24
& 1.11, -1.14
& 2.34, -1.61 \\
\bottomrule
\end{tabular}
\end{table*}

\begin{table*}[t]
\caption{\label{tab:si_atomic_coordinates}
Atomic coordinates and equivalent isotropic displacement parameters for
the reported $M$-Pb-Bi structures.}
\centering
\begin{tabular}{llccccc}
\toprule
Composition & Atom & Occ. & $x$ & $y$ & $z$ & $U_{\mathrm{eq}}/U_{\mathrm{iso}}$ (\AA$^2$) \\
\midrule
Pd--Pb--Bi & Bi1 & 0.5 & 0.29257(4) & 0.70743(4) & -0.5 & 0.02290(15) \\
& Pb1 & 0.5 & 0.29257(4) & 0.70743(4) & -0.5 & 0.02290(15) \\
& Bi2 & 0.5 & 0.60030(4) & 0.60030(4) & 0.5 & 0.02271(14) \\
& Pb2 & 0.5 & 0.60030(4) & 0.60030(4) & 0.5 & 0.02271(14) \\
& Pb3 & 0.5 & 0.49914(4) & 0.82961(5) & 0 & 0.02945(17) \\
& Bi3 & 0.5 & 0.49914(4) & 0.82961(5) & 0 & 0.02945(17) \\
& Pd1 & 1 & 0.41267(8) & 0.58733(8) & 0 & 0.0230(3) \\
\midrule
Rh--Pb--Bi & Bi1 & 0.63 & 0.29303(3) & 0.70697(3) & 0.5 & 0.01679(12) \\
& Pb1 & 0.37 & 0.29303(3) & 0.70697(3) & 0.5 & 0.01679(12) \\
& Bi2 & 0.63 & 0.40290(3) & 0.40290(3) & -0.5 & 0.01673(12) \\
& Pb2 & 0.37 & 0.40290(3) & 0.40290(3) & -0.5 & 0.01673(12) \\
& Pb3 & 0.37 & 0.49567(3) & 0.82601(4) & 0 & 0.02371(13) \\
& Bi3 & 0.63 & 0.49567(3) & 0.82601(4) & 0 & 0.02371(13) \\
& Rh1 & 1 & 0.41278(6) & 0.58722(6) & 0 & 0.0170(2) \\
\midrule
Au--Pb--Bi & Bi1 & 0.33 & 0.29152(4) & 0.70848(4) & 0 & 0.02975(17) \\
& Pb1 & 0.67 & 0.29152(4) & 0.70848(4) & 0 & 0.02975(17) \\
& Pb2 & 0.67 & 0.39232(4) & 0.39232(4) & 0 & 0.02838(15) \\
& Bi2 & 0.33 & 0.39232(4) & 0.39232(4) & 0 & 0.02838(15) \\
& Pb3 & 0.67 & -0.00116(5) & 0.33415(5) & 0 & 0.03503(17) \\
& Bi3 & 0.33 & -0.00116(5) & 0.33415(5) & 0 & 0.03503(17) \\
& Au1 & 1 & 0.08592(5) & 0.08592(5) & 0 & 0.03076(17) \\
\midrule
Ni--Pb--Bi & Bi1 & 0.6 & 0.08654(4) & -0.25 & 0.68074(3) & 0.01774(14) \\
& Pb1 & 0.4 & 0.08654(4) & -0.25 & 0.68074(3) & 0.01774(14) \\
& Bi2 & 0.6 & 0.21046(4) & 0.75 & 0.39692(3) & 0.01794(14) \\
& Pb2 & 0.4 & 0.21046(4) & 0.75 & 0.39692(3) & 0.01794(14) \\
& Pb3 & 0.4 & 0.38324(5) & 0.25 & 0.59912(3) & 0.02687(16) \\
& Bi3 & 0.6 & 0.38324(5) & 0.25 & 0.59912(3) & 0.02687(16) \\
& Ni1 & 1 & 0.08745(13) & 0.25 & 0.52078(11) & 0.0150(3) \\
\bottomrule
\end{tabular}
\end{table*}

\begin{table*}[t]
\caption{\label{tab:si_adps}
Anisotropic displacement parameters for the reported $M$-Pb-Bi structures.
All values are given in \AA$^2$.}
\centering
\begin{tabular}{llcccccc}
\toprule
Composition & Atom & $U_{11}$ & $U_{22}$ & $U_{33}$ & $U_{12}$ & $U_{13}$ & $U_{23}$ \\
\midrule
Pd--Pb--Bi & Bi1 & 0.0232(2) & 0.0232(2) & 0.0224(3) & 0.0027(2) & 0 & 0 \\
& Pb1 & 0.0232(2) & 0.0232(2) & 0.0224(3) & 0.0027(2) & 0 & 0 \\
& Bi2 & 0.0236(2) & 0.0236(2) & 0.0210(3) & -0.0012(2) & 0 & 0 \\
& Pb2 & 0.0236(2) & 0.0236(2) & 0.0210(3) & -0.0012(2) & 0 & 0 \\
& Pb3 & 0.0280(3) & 0.0250(3) & 0.0353(3) & -0.00138(16) & 0 & 0 \\
& Bi3 & 0.0280(3) & 0.0250(3) & 0.0353(3) & -0.00138(16) & 0 & 0 \\
& Pd1 & 0.0231(4) & 0.0231(4) & 0.0228(5) & 0.0049(4) & 0 & 0 \\
\midrule
Rh--Pb--Bi & Bi1 & 0.01627(18) & 0.01627(18) & 0.0178(3) & 0.00226(14) & 0 & 0 \\
& Pb1 & 0.01627(18) & 0.01627(18) & 0.0178(3) & 0.00226(14) & 0 & 0 \\
& Bi2 & 0.01649(17) & 0.01649(17) & 0.0172(3) & -0.00178(14) & 0 & 0 \\
& Pb2 & 0.01649(17) & 0.01649(17) & 0.0172(3) & -0.00178(14) & 0 & 0 \\
& Pb3 & 0.0225(2) & 0.0185(2) & 0.0301(3) & -0.00125(12) & 0 & 0 \\
& Bi3 & 0.0225(2) & 0.0185(2) & 0.0301(3) & -0.00125(12) & 0 & 0 \\
& Rh1 & 0.0170(3) & 0.0170(3) & 0.0170(5) & 0.0037(3) & 0 & 0 \\
\midrule
Au--Pb--Bi & Bi1 & 0.0275(2) & 0.0275(2) & 0.0343(4) & 0.0029(3) & 0 & 0 \\
& Pb1 & 0.0275(2) & 0.0275(2) & 0.0343(4) & 0.0029(3) & 0 & 0 \\
& Pb2 & 0.0261(2) & 0.0261(2) & 0.0328(3) & -0.0028(2) & 0 & 0 \\
& Bi2 & 0.0261(2) & 0.0261(2) & 0.0328(3) & -0.0028(2) & 0 & 0 \\
& Pb3 & 0.0330(3) & 0.0293(3) & 0.0428(3) & 0.00162(19) & 0 & 0 \\
& Bi3 & 0.0330(3) & 0.0293(3) & 0.0428(3) & 0.00162(19) & 0 & 0 \\
& Au1 & 0.0287(2) & 0.0287(2) & 0.0350(4) & -0.0052(3) & 0 & 0 \\
\midrule
Ni--Pb--Bi & Bi1 & 0.0220(3) & 0.0160(3) & 0.0152(2) & 0 & -0.00298(11) & 0 \\
& Pb1 & 0.0220(3) & 0.0160(3) & 0.0152(2) & 0 & -0.00298(11) & 0 \\
& Bi2 & 0.0181(3) & 0.0165(3) & 0.0192(2) & 0 & 0.00505(11) & 0 \\
& Pb2 & 0.0181(3) & 0.0165(3) & 0.0192(2) & 0 & 0.00505(11) & 0 \\
& Pb3 & 0.0162(2) & 0.0405(3) & 0.0239(3) & 0 & -0.00402(13) & 0 \\
& Bi3 & 0.0162(2) & 0.0405(3) & 0.0239(3) & 0 & -0.00402(13) & 0 \\
& Ni1 & 0.0151(6) & 0.0130(6) & 0.0169(6) & 0 & -0.0011(4) & 0 \\
\bottomrule
\end{tabular}
\end{table*}

% ---------------------------------------------------------------------
% Property figures (Figures S5-S10)
% ---------------------------------------------------------------------
\begin{figure}[h]
\centering
\includegraphics[width=\textwidth]{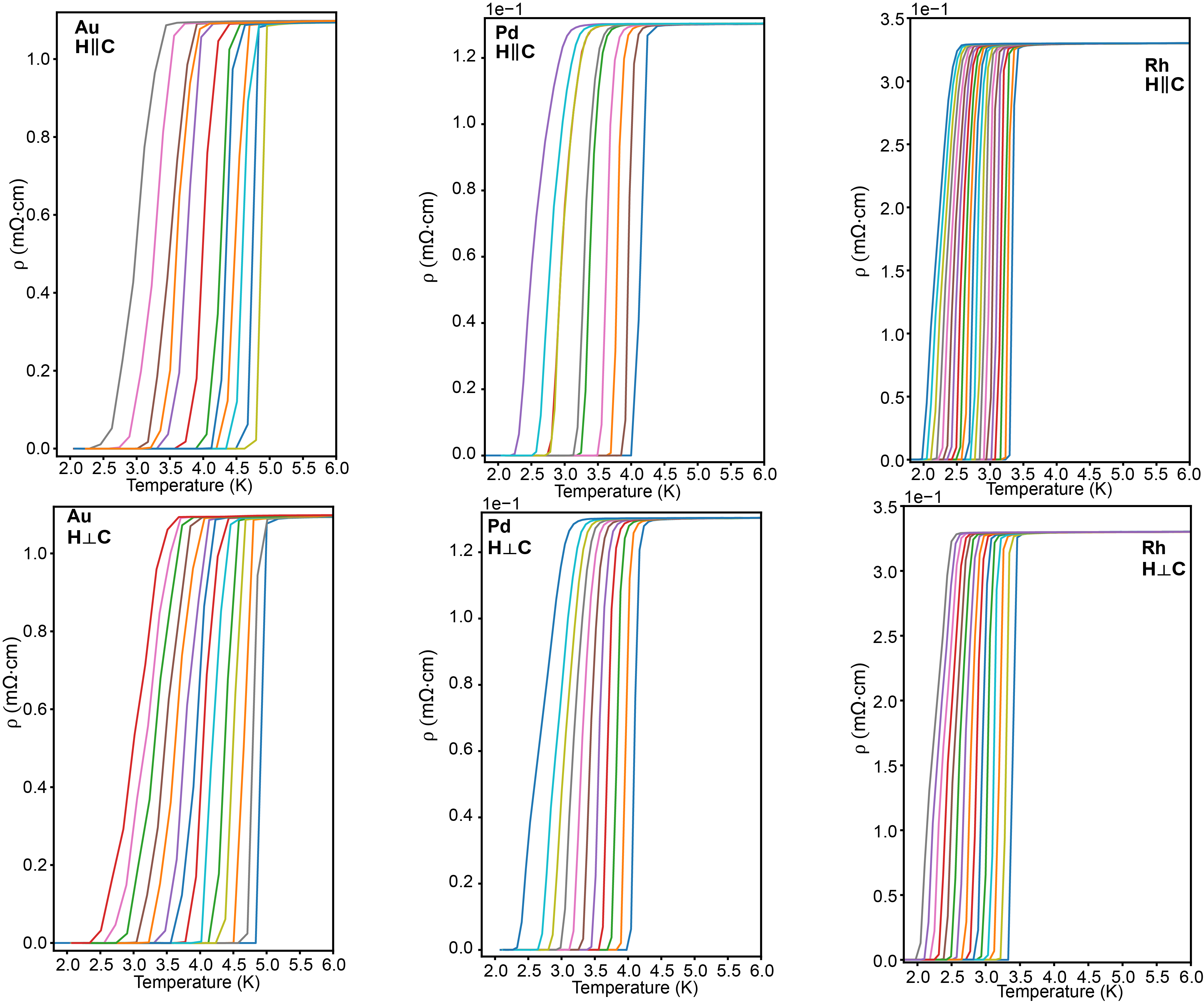}
\caption{Raw resistivity under increasing magnetic field, parallel and
perpendicular to the crystallographic $c$ axis.}
\label{fig:S5}
\end{figure}

\begin{figure}[h]
\centering
\includegraphics[width=\textwidth]{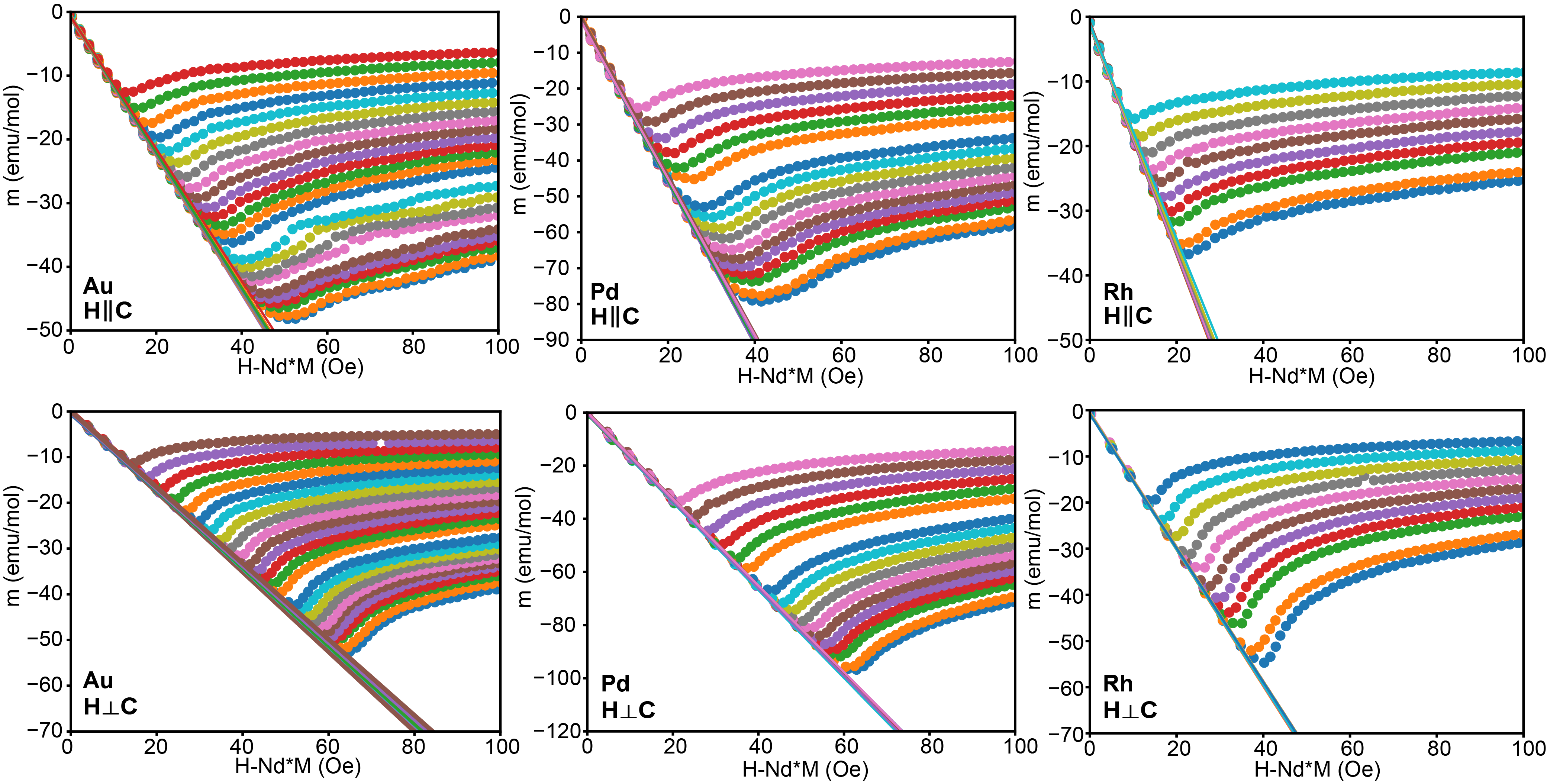}
\caption{Raw magnetic moment versus magnetic field with increasing
temperature. Lines are linear fits to the Meissner curve.}
\label{fig:S6}
\end{figure}

\begin{figure}[h]
\centering
\includegraphics[width=0.6\textwidth]{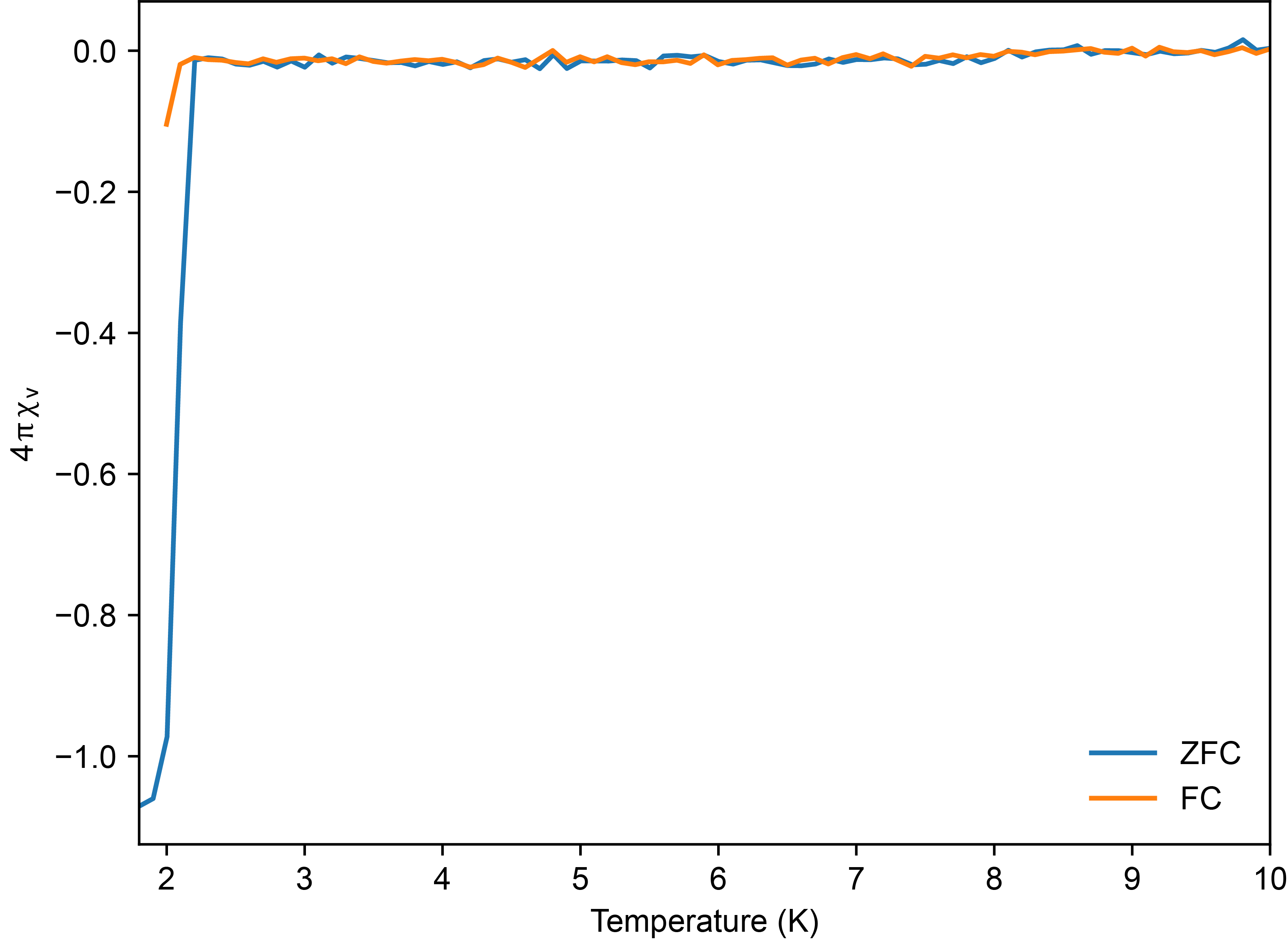}
\caption{Demagnetization-corrected ZFC and FC magnetic susceptibility of
NiBi$_{3-x}$Pb$_x$ under a 2~Oe bias field.}
\label{fig:S7}
\end{figure}

\begin{figure}[h]
\centering
\includegraphics[width=0.6\textwidth]{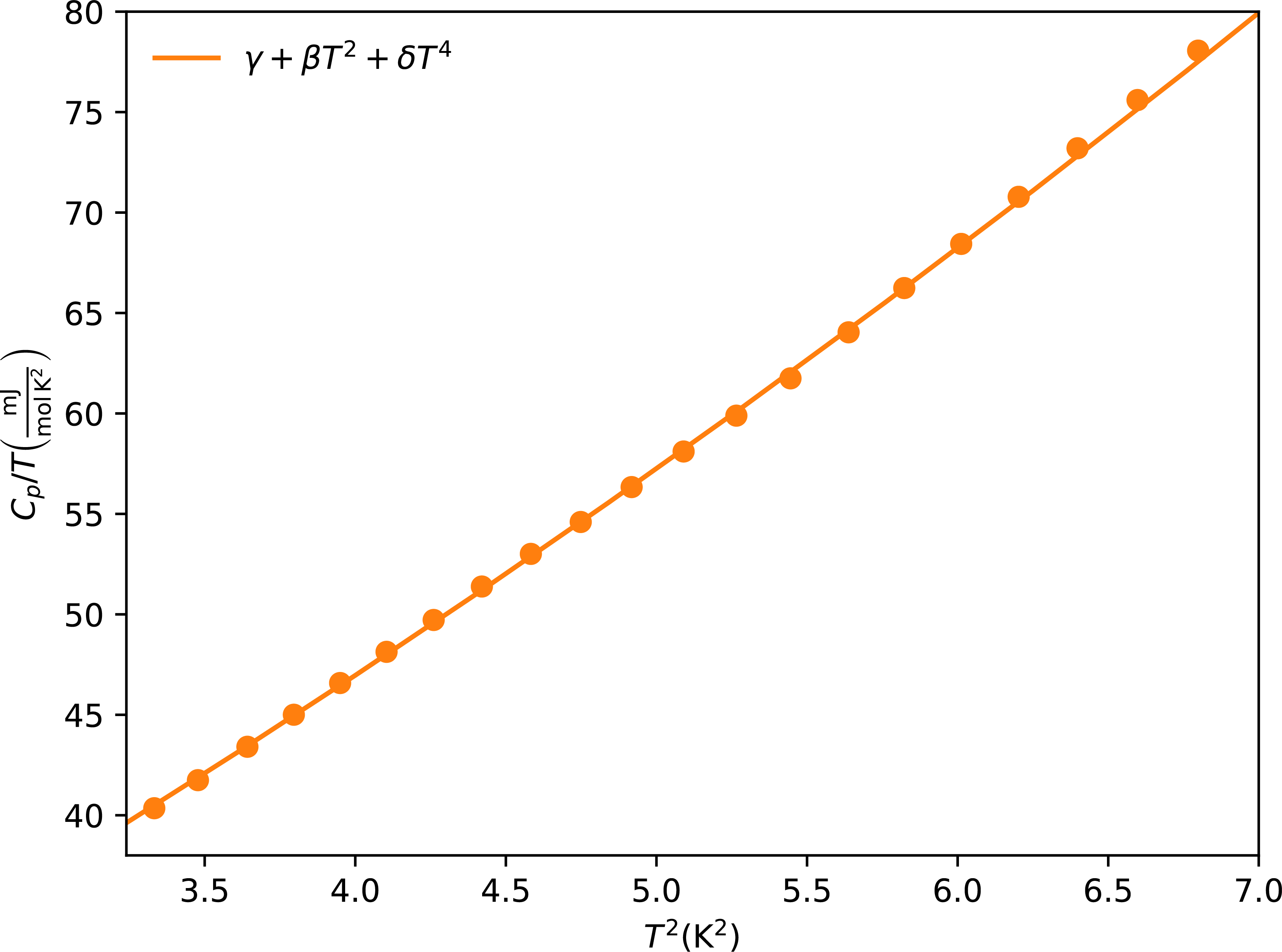}
\caption{Low-temperature measurement and fit of the total specific heat
$C_p/T$ for AuPb$_{4-x}$Bi$_x$ in a 50~kOe field parallel to the
crystallographic $ab$ plane.}
\label{fig:S8}
\end{figure}

\begin{figure}[h]
\centering
\includegraphics[width=0.6\textwidth]{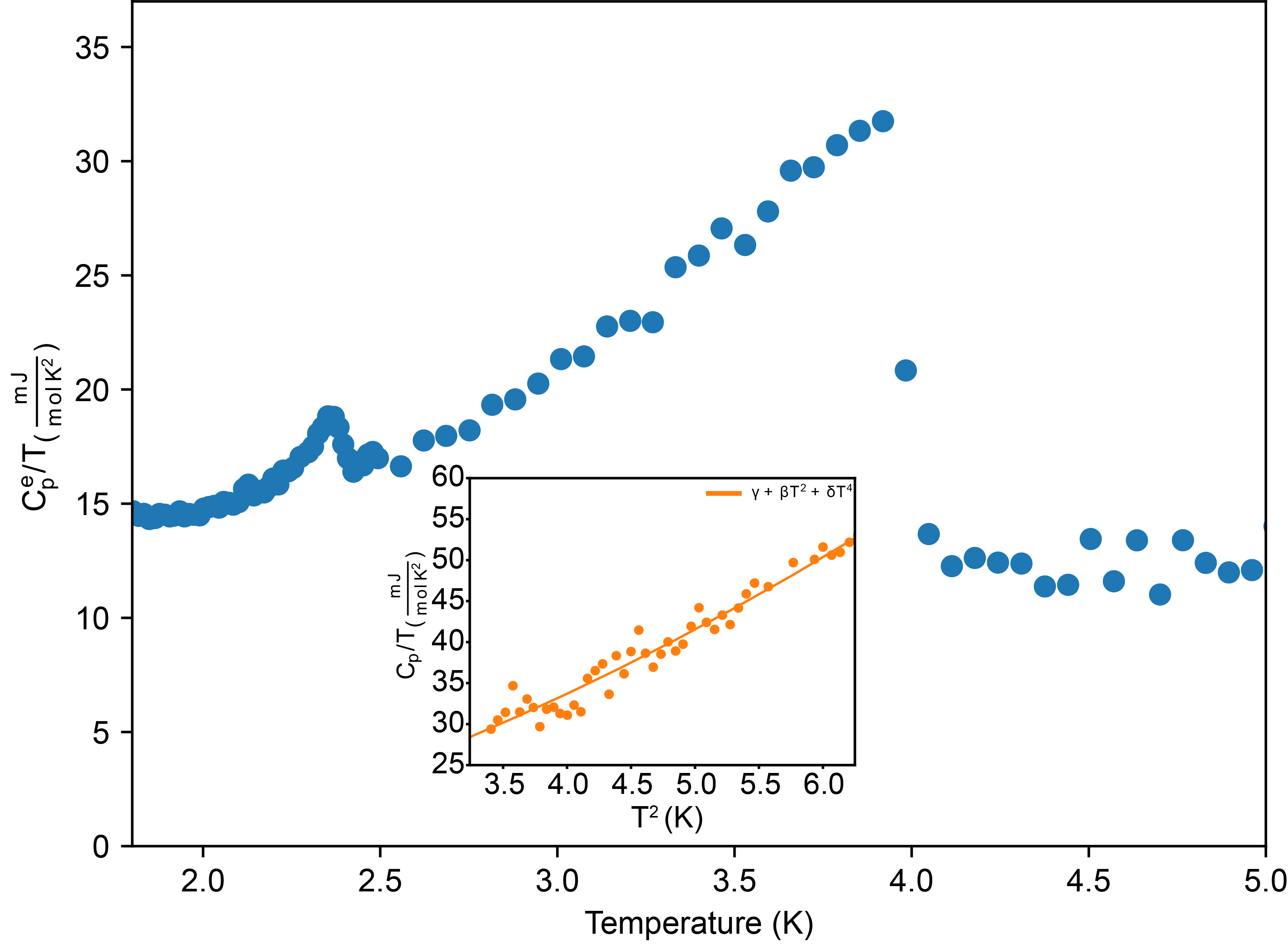}
\caption{Electronic specific heat $C_p^e/T$ of PdPb$_{4-x}$Bi$_x$ in zero
field. An unidentified transition is observed also at $\approx2.4$~K. Inset: low-temperature measurement and fit of the total specific
heat $C_p/T$ in a 50~kOe field parallel to the crystallographic $ab$ plane.}
\label{fig:S9}
\end{figure}

\begin{figure}[h]
\centering
\includegraphics[width=0.6\textwidth]{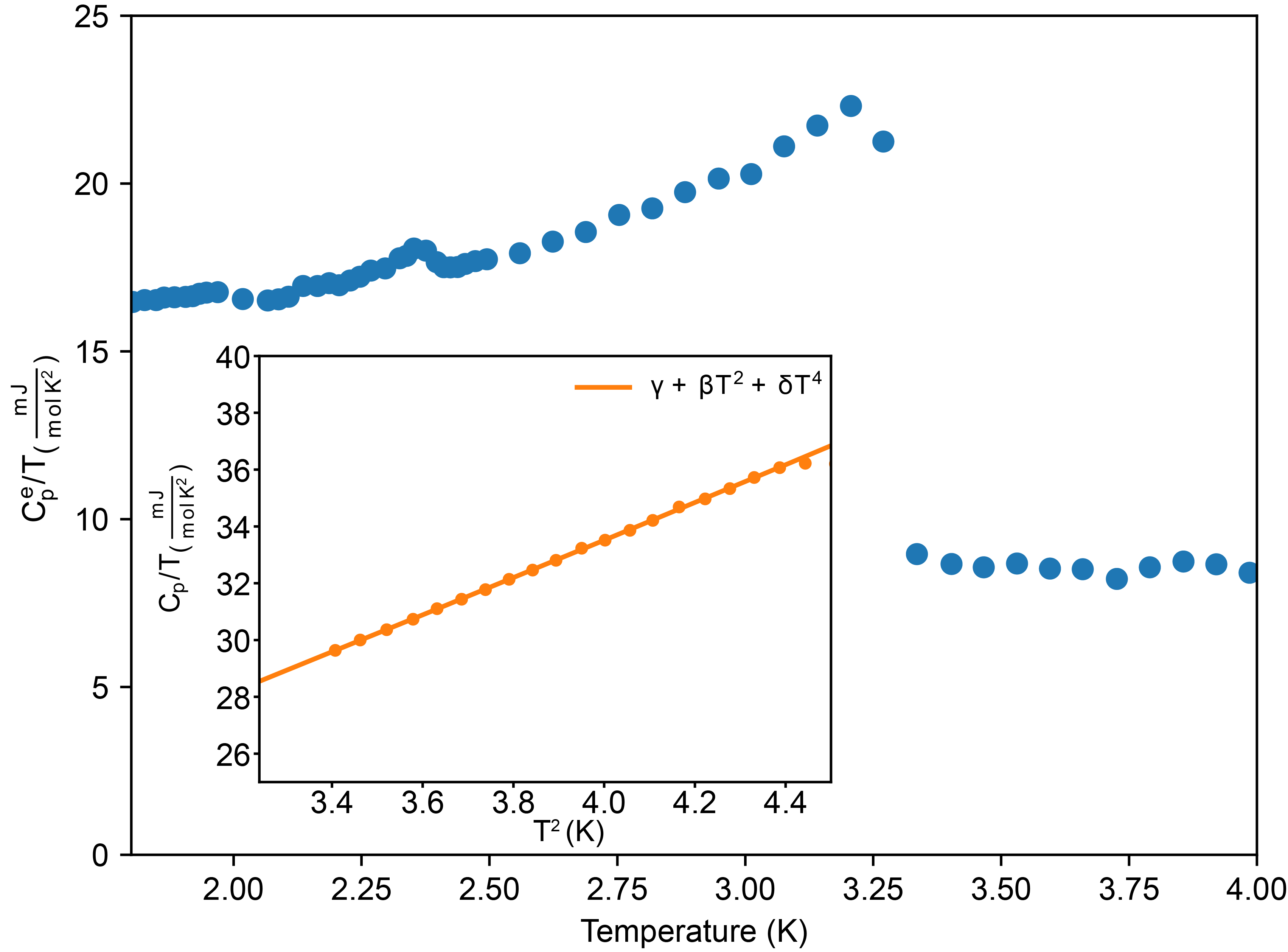}
\caption{Electronic specific heat $C_p^e/T$ of RhPb$_{4-x}$Bi$_x$ in zero
field. An unidentified transition is observed also at $\approx2.4$~K. Inset: low-temperature measurement and fit of the total specific
heat $C_p/T$ in a 50~kOe field parallel to the crystallographic $ab$ plane.}
\label{fig:S10}
\end{figure}

\begin{figure}[h]
\centering
\includegraphics[width=0.8\textwidth]{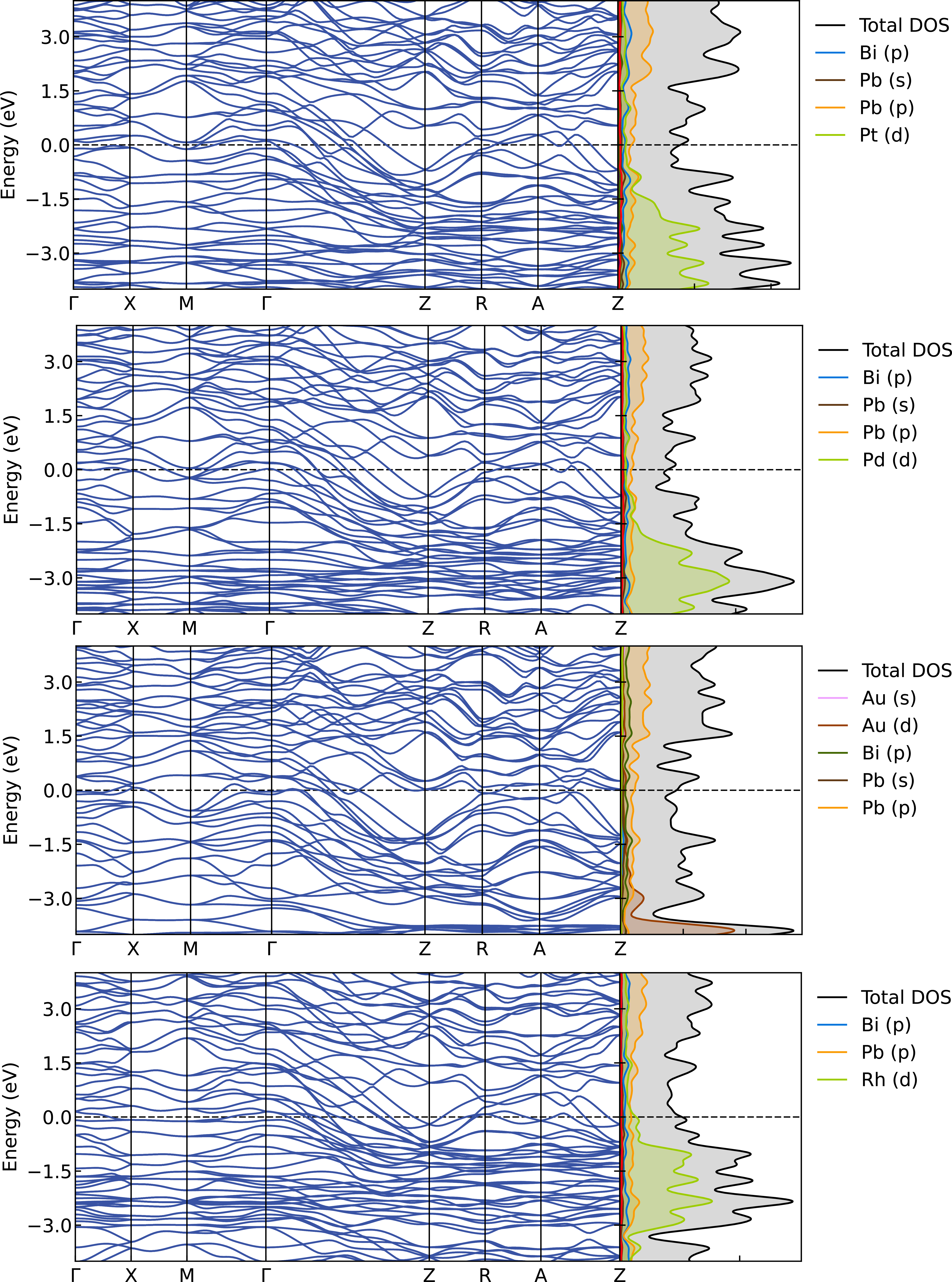}
\caption{Calculated band structures and total and partial density of states for the $M$Pb$_{4-x}$Bi structures}
\label{fig:S11}
\end{figure}

\end{document}